\documentclass[manuscript,nonacm]{acmart}
\usepackage{longtable}
\usepackage{array}
\AtBeginDocument{%
 }

\setcopyright{acmlicensed}
\copyrightyear{2024}
\acmYear{2024}
\acmDOI{XXXXXXX.XXXXXXX}

\begin{document}

\title{Global Overview of Computational Thinking and Digital Tools for Teaching}

\author{Roberto Massi de Oliveira}
\affiliation{%
 \institution{UNICAMP}
 \city{Campinas}
 \state{SP}
 \country{Brazil}}
\email{rmo@unicamp.br}

\author{Mônica Cristina Garbin}
\affiliation{%
 \institution{UNICAMP}
 \city{Campinas}
 \state{SP}
 \country{Brazil}}
\email{mgarbin@unicamp.br}

\author{Rodolfo Azevedo}
\affiliation{%
 \institution{UNICAMP}
 \city{Campinas}
 \state{SP}
 \country{Brazil}}
\email{rodolfo.azevedo@unicamp.br}

\begin{abstract}
Computational Thinking (CT) has emerged as a critical component in modern education, essential to equip students with the skills necessary to thrive in a technology-driven world. This survey provides a comprehensive analysis of the presence and integration of CT in school curricula across various countries. In addition, this study categorizes digital tools into groups such as visual programming, textual programming, electronic games, modeling, and simulation, assessing their use in different educational settings. Furthermore, it examines how these tools are employed in various contexts, including the areas of knowledge and age groups they target, and the specific skills they help develop. The research also identifies key CT competencies that have been improved through these tools, including Cognitive and Analytical Competencies (CAC), Technical and Computational Competencies (TCC) and Social and Emotional Competencies (SEC). Furthermore, the study highlights recurring challenges in the implementation of digital tools for CT development, such as inadequate infrastructure, difficulties in the usability of the tool, teacher training, adapting pedagogical practices, and measuring student CT skills. Finally, it proposes areas for future research to address these challenges and advance CT education. 
\end{abstract}



\received{XX August XX}
\received[revised]{XX August XX}
\received[accepted]{XX August XX}

\maketitle

\section{Introduction}

The increasing importance of Computational Thinking (CT) in education is driven by its critical role in equipping students with the essential problem solving and analytical skills necessary for the digital age. CT involves solving problems, designing systems, and understanding human behavior using concepts fundamental to computer science. The potential of digital tools to facilitate CT development is particularly promising, as they provide interactive and engaging platforms for students to learn and apply these skills in various contexts~\cite{arMonjelat201962, arSilva202291}.

Globally, the emphasis on CT in education has grown as countries recognize the need to prepare students for a digital future. Various countries have incorporated CT into their school curricula, either as standalone subjects or as competencies within broader areas such as computer science or information and communication technologies (ICT). This trend reflects a broader understanding of the importance of digital literacy and the ability to navigate complex technological environments~\cite{arTran2019100}. Countries such as the United States, the United Kingdom and several European nations have been at the forefront of this movement, integrating CT into national education strategies. Meanwhile, emerging economies are also making strides, albeit at different paces, in incorporating these skills into their education systems.

Despite the global push towards CT education, several challenges and gaps remain. A significant issue is the disparity in access to technology and teacher training~\cite{arSaez-Lopez202085}, which affects the implementation and effectiveness of CT education, especially in under-resourced regions. Furthermore, there is a lack of consensus on the most effective methods for teaching CT, as well as on the assessment of CT skills~\cite{arTang202097}. The varying levels of digital infrastructure, cultural differences in educational priorities, and the need for localized curriculum adaptations further complicate the global landscape of CT education~\cite{arEryilmaz202124, arGokce202332}. These challenges highlight the need for ongoing research and innovation in teaching methods and tools to bridge the gaps in CT education.

The primary objective of this article is to investigate and organize the state-of-the-art in the integration of CT into school curricula and the use of digital tools for teaching CT on a global scale. Specific objectives include analyzing the inclusion of CT in school curricula in different countries, identifying and classifying the policies and strategies used to integrate CT, categorizing the digital tools used in teaching CT, examining the use of these tools in various educational contexts, and identifying the CT competencies developed through these tools. The study also aims to identify challenges and gaps in the implementation of CT education and to provide insight into potential areas for future research and innovation in this field.

The remainder of this article is structured as follows. Section~\ref{secRelated} reviews related work and surveys on CT education, highlighting key findings and trends. Section~\ref{secMethodology} outlines the methodology used for the review of the literature and data analysis. Section~\ref{secOverview} presents a global overview of the integration of CT into school curricula, detailing the policies and strategies adopted by different regions. Section~\ref{secTools} discusses the classification and evaluation of various digital tools used in the teaching of CT. Section~\ref{secOpportunities} identifies the competencies developed through these tools and addresses the challenges and gaps in CT education. Finally, Section~\ref{secConcluding} presents concluding remarks.

\section{Related Surveys} \label{secRelated}

The related surveys converge to highlight the importance of CT in education, using digital tools in various educational contexts, from early childhood education to higher education. Systematic reviews and meta-analyses emphasize the effectiveness of these tools in improving cognitive skills, problem solving abilities, critical thinking, creativity, and collaboration. Furthermore, all related surveys, as well as the present work, share the common goal of encouraging the academic community to produce new research to fill the identified gaps, promoting advances in the teaching of computational thinking, and improving the implementation of digital tools in educational contexts.

In~\cite{arZeng2023119}, the authors conduct a systematic review of the literature, with the main contributions including the adaptation of~\citet{inBrennan2012}'s CT framework for the context of early childhood education, identification and definition of appropriate CT components for children aged 2 to 8, and the proposal of a new CT curriculum framework that includes concepts, practices, and perspectives. In addition, the article provides theoretical and practical guidelines for educators and policy makers on how to incorporate CT into early childhood education, highlighting the importance of structured CT instruction from an early age.

Several studies have explored the impact of block-based visual programming tools, mainly Scratch. For example, in~\cite{arEstevez201925}, the Scratch digital tool was used as an integral part of the study, aimed at high school students aged 16 to 17. The main contribution of the study lies in the innovative approach to introducing fundamental concepts of Artificial Intelligence (AI) and CT through an experiential learning framework. The main digital tools discussed in~\cite{arSun202295} were Scratch, Alice, and educational robotics. The results indicated that programming ability includes metacognition, cognition, operation, and communication skills, with a greater focus on operational ability. In~\cite{arHu202139}, the authors examined the use of Scratch, Alice, and MIT App Inventor in the learning process of programming for novice students at different educational levels. The results highlighted significant improvements in student performance, especially for those in elementary and high school, with Scratch being more effective in promoting academic success compared to Alice and MIT App Inventor.

Many other surveys explored Scratch functionalities that support the development of CT skills, mostly within the K-12 range~\cite{arZhang2019120, arFagerlund202127, arIbrohim202340, arMontiel202164}, in a variety of knowledge fields. For example, in~\cite{arDuo-Terron202320}, the use of Scratch was analyzed in the educational context to develop CT and STEAM skills (science, technology, engineering, art and mathematics). In~\cite{arFidai202029}, the research aimed to calculate the overall effect of Arduino and Scratch-based interventions in STEM (science, technology, engineering and mathematics) education. 

In~\cite{arPerez-Jorge202277}, the authors used Scratch and App Inventor digital tools, targeting university students from various fields, including computing, engineering, programming, and activities focused on gamification, interculturality, multimedia, mathematics, and psychology. The results indicated that the use of these tools contributed positively to the development of students' skills and competencies, including autonomy, attention, motivation, critical thinking, creative thinking, communication, problem solving, and social interaction.

The review~\cite{arRich202280} involved cataloging more than 300 tools that are intended to teach programming to children. The research covered more than 150 studies that implemented CT interventions in elementary school students from 35 countries and was available in 8 different languages. The most common tools used in the studies included Scratch, with a prevalence of 35\%, followed by a wide variety of other tools such as MakeCode, Bee-bots, micro, and e-textiles, reflecting the diversity of options available to educators. The conclusions highlight the rapid growth of research on CT teaching in primary education, highlighting the importance of the accessibility and variety of tools available to educators.

In~\cite{arKuz202250}, the authors aim to analyze and review different educational software for learning programming in playful environments, including Scratch, Lightbot, PilasEngine, Pilas Bloques, Turtle Art, and Blockly. As a result, the study presents a detailed comparison of these tools, highlighting their advantages and disadvantages in teaching programming. The conclusions point out that these educational software strengthen cognitive and problem-solving skills in students and that the choice of tool should consider the educational context and pedagogical objectives.

In the subject of teacher training, in~\cite{arKong202347}, the authors investigated the effectiveness of a scalable and sustainable professional development program for teachers when teaching CT (CT) in primary schools in Hong Kong. The conclusions highlight the importance of continuing support for teachers in implementing CT initiatives, especially with regard to the diversity of students and the integration of CT with teaching other subjects. In~\cite{arErsozlu202323}, the authors explored the capacity and applicability of the methods, tools, and approaches used to teach CT through mathematics, also analyzing the professional development of teachers. Similarly, in~\cite{arSantos202389}, the authors investigated the training of mathematics teachers focusing on integrating digital technologies such as Scratch, App Inventor, Thunkable, Tinkercad, and Arduino in Basic Education. The study aims to understand the role of these technologies in Mathematics Education, identify the challenges faced by teachers, and propose strategies to improve their training. The results indicate that integrating these technologies can make Mathematics teaching more engaging and practical, but teachers face challenges such as acquiring technological skills, adapting pedagogical strategies, and managing time and resources. In addition, they emphasize the need for institutional support and the promotion of a culture of innovation in schools to overcome resistance to change and improve teaching practices.

Finally, in~\cite{arTang202097}, the article systematically reviews how CT skills have been assessed in the literature. The results indicate the need for more CT assessments for high school students, university students, and professional development programs for teachers. It concludes that most CT assessments focus on students' programming or computing skills, with a high proportion of studies conducted in formal education contexts.

As can be seen in analyzing the aforementioned research, there is considerable progress in the state of the art concerning the use and benefits of digital tools in teaching CT. However, unlike previous work, this research presents the following contributions in an articulated manner.

\begin{itemize}
    \item \textbf{Knowledge of the Global Landscape of CT Education:} we provide readers with a global overview of the inclusion of CT in school curricula.
    \item \textbf{Comprehensive Understanding of Digital Tools:} we offer readers a clear view of the various digital tools available for teaching CT, helping in the selection of an appropriate tool for a specific educational context.
    \item \textbf{Understanding of Developed Competencies:} we provide readers with a clearer and more organized view of CT skills that have already been enhanced with the help of these tools.
\end{itemize}

\section{Methodology} \label{secMethodology}

The general objective of this work is to investigate and organize the state-of-the-art in the use of digital tools to teach CT in a global context. To this end, the authors initially identified search terms related to the main objective of the study. These keywords were then used to perform searches on the indexers, with the total occurrences for each keyword shown in Table~\ref{tab:keywords}. The first phase of the literature review took place in December 2023 and was repeated in May 2024, including new works.

The research was carried out using the CAPES Portal of Journals, which is an aggregator, offering a wide range of technical-scientific content, including collections of journals, libraries, digital repositories, and national and international databases, totaling 341 databases for search.

\begin{table}[h]
\centering
\caption{Distribution of keywords and corresponding articles.}
\label{tab:keywords}
\begin{tabular}{|l|c|c|c|}
\hline
\textbf{Keywords} & \textbf{Search} & \textbf{Abstract/ Title} & \textbf{Full Reading} \\ \hline
Computational Thinking + Survey & 153 & 10 & 9 \\ \hline
Computational Thinking + Tools & 629 & 5 & 4 \\ \hline
Computational Thinking + Tools + Review of literature & 51 & 9 & 5 \\ \hline
Computational Thinking + Scratch & 207 & 66 & 63 \\ \hline
Computational Thinking + MIT App Inventor & 8 & 6 & 6 \\ \hline
Blockly & 30 & 4 & 4 \\ \hline
Python Tutor & 9 & 1 & 1 \\ \hline
Computational Thinking + Alice & 6 & 3 & 3 \\ \hline
Google Colab + Programming & 31 & 1 & 1 \\ \hline
Tinkercad & 37 & 1 & 1 \\ \hline
Blockly Games & 3 & 2 & 2 \\ \hline
LightBot & 7 & 6 & 6 \\ \hline
Computational Thinking + Malt2 & 2 & 2 & 2 \\ \hline
Computational Thinking + Zoombinis & 3 & 2 & 2 \\ \hline
Computational Thinking + Code.org & 15 & 7 & 6 \\ \hline
Computational Thinking + Swift Playgrounds & 1 & 1 & 1 \\ \hline
Computational Thinking + Logo & 36 & 2 & 2 \\ \hline
Computational Thinking + Pencil Code & 5 & 1 & 1 \\ \hline
Computational Thinking + MyClassGame & 1 & 1 & 1 \\ \hline
\textbf{Total} & \textbf{1234} & \textbf{130} & \textbf{120} \\ \hline
\end{tabular}
\end{table}

\begin{figure}[h]
\centering
\includegraphics[width=0.75\linewidth]{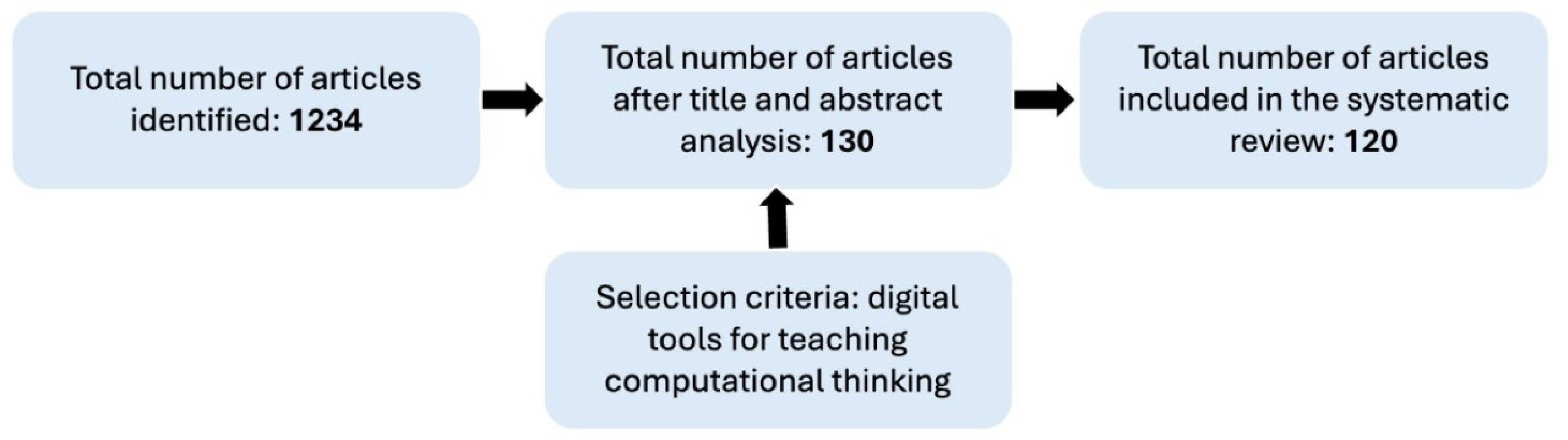}
\Description{Flowchart showing the filtering of articles. The process starts with 1234 identified articles, narrows down to 130 articles after title and abstract analysis, and finally includes 120 articles in the systematic review. There is a criterion for selection: digital tools for developing computational thinking.}
\caption{Methodology steps to select the articles}
\label{fig:artigos}
\end{figure}

The diagram in Figure~\ref{fig:artigos} illustrates the filtering steps and the specific criteria applied to ensure the relevance and quality of the articles in the systematic review. Initially, 1234 articles were identified based on the 19 search terms listed in Table~\ref{tab:keywords}. After an analysis of titles and abstracts, this number was reduced to 130. Finally, 120 articles were included in the systematic review.

The selection criteria included the use of digital tools for CT development and the publication of articles in journals between 2019 and 2024. Therefore, works that did not exclusively address digital tools for the development of CT were excluded, such as articles on robotics that did not cover CT and required the use of physical boards, or studies focusing on unplugged computing.

After analyzing the studies, the authors, based on the objectives of the work, sought to identify the status of CT in the school curricula of the 34 countries whose publications were selected for this work. These countries are: Argentina, Australia, Brazil, Canada, Kazakhstan, Czech Republic, Chile, China, Colombia, South Korea, Croatia, Ecuador, Spain, United States, Finland, Greece, Indonesia, Ireland, Israel, Italy, Jordan, Malaysia, Malta, Morocco, Mexico, Norway, Paraguay, Peru, Poland, Portugal, Sweden, Thailand, Taiwan, and Turkey.

To carry out this survey, the search was conducted in official documents of the Ministry of Education or equivalent institutions. Search tools on the Internet and artificial intelligence tools were utilized to assist in identifying documents from official channels. In addition, online translation tools were employed to interpret documentation written in languages unfamiliar to the authors of this study.

For each country, a law, decree, model, dispatch, or equivalent document related to the official curriculum of elementary education schools was identified. During the review of these documents, terms such as "Computer Science," "ICT competence," and "Computational Thinking" were specifically searched for.

\section{Computational Thinking and Global Overview} \label{secOverview}

The introduction of technology as a subject in basic education reflects the rapid evolution of society and the growing importance of technological skills in the contemporary world. This process began to gain momentum in the second half of the twentieth century when computers became more accessible, and schools started recognizing the importance of preparing students for a digital future. The initial focus was primarily on developing basic computer handling skills, such as text applications, spreadsheets, and later, internet navigation. Early initiatives were often limited by resource availability and the need for teacher training.

It was in the 1980s that the teaching of programming became clearer with the Logo language, developed by Seymour Papert~\cite{arSanchezCamacho202387}, who argued that its learning could help students develop critical and creative thinking skills. However, it was from 2006 onward that the concept of CT began to be popularized, with the famous article by Jeannette Wing titled "Computational Thinking," published in March 2006 in the Communications of the ACM (Association for Computing Machinery) journal. In this article, Wing argues that CT is a fundamental competence that should be taught to everyone, not just computer scientists. She defines CT as a process that "involves solving problems, designing systems, and understanding human behavior, by drawing on the concepts fundamental
to computer science. Computational thinking includes a range of mental tools that reflect the breadth of the field of computer science" (p. 33). This article is widely recognized for bringing the concept of CT to the forefront of education and research in Computer Science.

Similarly, other authors \cite{arHsu202238, arDiaz-Lauzurica201919, arKarakasis202045, arELOY202122, arBroza202309, arPerez-Marin202078, arPapavlasopoulou201973, arSanchezCamacho202387} define CT as the process of thinking that involves solving problems, designing systems, and understanding human behavior based on fundamental concepts of computer science. Furthermore, it is described as a methodology that uses different levels of abstraction, analytical and algorithmic approaches to formulate, analyze, and solve problems \cite{arUrquizo2021106, arDuo-Terron202320, arErsozlu202323, arKourti202348, arMonjelat201962, arPiedade202279}. CT is also associated with the ability to solve problems using automated techniques by recognizing and abstracting problems based on basic concepts and principles of computational technology \cite{arKYNIGOS201851, arCalandra202111, arHsu202137, arRochadiani202381, arAttard202005}.

In expanding its conception, many authors draw on elements of Computer Science. In~\cite{arKuz202250}, CT is understood as a problem solving process that includes problem formulation, data collection, representation, and analysis through abstractions such as models and simulations, automated solutions through algorithmic thinking, identification, analysis, and implementation of possible solutions that aim for the most efficient and effective combination of resources and steps, and generalization and transfer of this process to a variety of contexts. In~\cite{arZampieri2020118}, CT is described as a problem solving process that includes formulating problems in a way that allows the use of computers and other tools to help solve them, organizing and analyzing data logically, representing data through abstractions such as models and simulations, automating solutions through algorithmic thinking, identifying, analyzing, and implementing possible solutions to achieve the most efficient and effective combination of steps and resources, and generalizing and transferring this problem solving process to a variety of problems.

Moreover, although the notion of the importance of adopting CT as a competence to be developed by students in basic education has expanded globally, the work~\cite{arMechelen202357} sought to map the state of the art of emerging technologies in education with the world's K-12 audience. During data analysis, the authors identified a rapid expansion of global interest in emerging technologies in basic education, with a particular emphasis on artificial intelligence and machine learning. However, a limited geographical distribution was observed in academic production, with most of the studies originating in Europe and North America. The main contributions of the study were the identification of gaps in understanding and implementing these technologies in schools, highlighting the need for a more integrated approach to school curricula, as well as the inclusion of ethical and social aspects related to the use of these technologies.

\begin{figure}[h]
\centering
\includegraphics[width=0.9\linewidth]{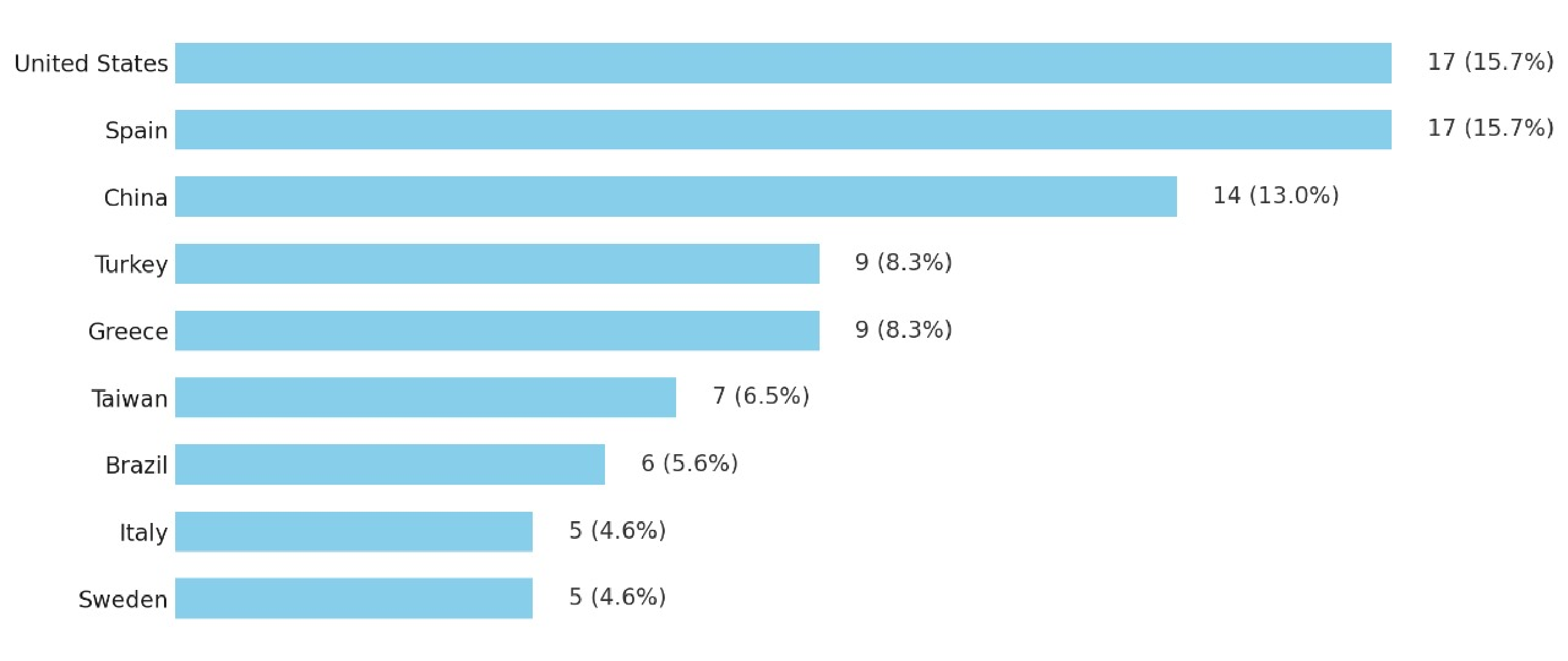}
\Description{Bar chart showing the number of computational thinking articles by country. The United States and Spain each have 17 articles (15.7\%), China has 14 articles (13.0\%), Turkey and Greece each have 9 articles (8.3\%), Taiwan has 7 articles (6.5\%), Brazil has 6 articles (5.6\%), Italy and Sweden each have 5 articles (4.6\%).}
\caption{Number of computational thinking articles by country, considering countries with at least 5 publications}
\label{fig:paises}
\end{figure}

In this sense, Figure~\ref{fig:paises} confirms the data presented by the authors, as among the 120 articles analyzed between 2019 and 2024 for this work, a significant concentration of publications can be observed in the United States and European countries. However, it is also possible to verify the inclusion of Asian countries such as China and Taiwan, as well as South American countries such as Brazil.

The adoption of CT in some countries faces significant challenges, often related to socioeconomic conditions. In nations where educational resources are limited and access to technology is restricted, implementing curricula that incorporate computational skills becomes more difficult. The lack of adequate infrastructure and the shortage of trained teachers to teach these competencies exacerbate the problem. Furthermore, in contexts where educational priorities focus on basic needs, the inclusion of new subjects may be perceived as less urgent. This socioeconomic disparity prevents many students from acquiring essential 21st century skills, exacerbating educational inequality, and limiting their future opportunities.

However, despite these inequalities, the adoption of CT has been expanding worldwide and is incorporated into school curricula and public policies in various countries. In addition to the countries described in Figure~\ref{fig:paises}, Table~\ref{tab:curricula} shows the scenario of the 34 countries with publications between 2019 and 2024 involving the teaching of CT mediated by digital technologies in basic education.

The documents consulted are official and issued by the Ministry of Education or a similar entity in the country, with the exception of three countries. Canada, the United States, and Malaysia. This approach was chosen to provide a broader view of CT understanding in the countries targeted by the research. It should be noted that official documents were consulted in their native language when an English version was not available. In such cases, the documentation was translated into the authors' language for subsequent reading.

In the case of Canada, the mandatory basic education curriculum is not a unified document, and each province is responsible for its documentation. However, it was possible to observe that there are six general competencies that guide the provincial documents: a. critical thinking and problem solving; b. innovation, creativity, and entrepreneurship; c. learning to learn/self-awareness and self-direction; d. collaboration; e. communication; f. global citizenship and sustainability. In the British Columbia curriculum, there is an area called "Applied Design, Skills, and Technologies curriculum," where we find the subarea "Information and Communications Technology." Although there is no specific mention of teaching CT, we find the content of "Computer Programming." In the Ontario curriculum, we find the areas of Science and Technology (K-8) and Computer Studies (9-12).

The United States also does not have a single, mandatory national curriculum for basic and secondary education. The responsibility for defining curricula lies with each state, school district, and even the school itself, based on their needs, resources, and priorities. The Computer Science Teachers Association (CSTA) created the K-12 Computer Science Standards and the curriculum for IT education.

Regarding Malaysia, the Prime Minister announced the integration of CT competencies in all subjects, starting in 2017 with primary and elementary school students. CT is listed as one of the components added to the revised Malaysian curriculum. The first step in integrating CT into the curriculum is to prepare teachers to apply the concepts in their daily teaching and learning practices. This information was obtained from~\cite{UNESCO2023}.

In addition to these three cases, we also highlight Morocco, where, in addition to the official document, information was also consulted in~\cite{Morocco2019}. Throughout Moroccan education, the introduction to modern information technologies, communication, and interactive creation is listed as skills to be developed. Since 2007, Computer Science has been a mandatory subject in secondary education, with students aged 11 to 16. In 2020/2021, computing was also introduced into primary education. Finally, the GENIE program is an operational implementation of the national strategy to generalize Information and Communication Technology in Education (ICTE) and is based on four main components: infrastructure, teacher training, digital resources, and improvement of practices.

\begin{longtable}{|p{1.5cm}|p{1.7cm}|p{5.3cm}|p{1cm}|p{1.5cm}|p{1.5cm}|}
\caption{Context of CT adoption and digital technology teaching in countries with publications (2019-2024) for digitally mediated teaching}
\label{tab:curricula} \\
\hline
\textbf{Country} & \textbf{Presence in School Curriculum} & \textbf{Consulted Document} & \textbf{Year} & \textbf{Language} & \textbf{References} \\ \hline
\endfirsthead
\caption[]{(Continued)} \\
\hline
\textbf{Country} & \textbf{Presence in School Curriculum} & \textbf{Consulted Document} & \textbf{Year} & \textbf{Language} & \textbf{References} \\ \hline
\endhead
\hline
\endfoot
Argentina & Competence & Resolution No. 1536-E/2017 & 2017 & Spanish & \cite{Argentina2017} \\ \hline
Australia & Competence & Australian Curriculum & 2013 & English & \cite{Australia2013} \\ \hline
Brazil & Competence & National Common Curricular Base (BNCC) & 2022 & Portuguese & \cite{Brazil2022} \\ \hline
Canada & NA & NA & NA & NA & NA \\ \hline
Kazakhstan & Competence & Digital Education Strategy and Implementation Plan & 2022 & English & \cite{Africa2022} \\ \hline
Czechia & Computer Science & Framework Educational Programme for Basic Education & 2007 & English & \cite{CzechRepublic2007} \\ \hline
Chile & Competence & Curricular Bases for 3rd and 4th year of secondary education (Supreme Decree N°193 of 2019) & 2020 & Spanish & \cite{Chile2020} \\ \hline
China & Computer Science & Notice from the Ministry of Education on the popularization of information technology education in primary and secondary schools & 2000 & Chinese & \cite{China2000} \\ \hline
Colombia & Competence & Curriculum Guidelines for the area of Technology and Informatics in Basic and Secondary Education & 2022 & Spanish & \cite{Colombia2022} \\ \hline
South Korea & Computer Science & The National Framework for the Elementary and Secondary Curriculum & 2022 & Korean & \cite{Korea2022} \\ \hline
Croatia & Computer Science & National Curriculum & 2018 & English & \cite{Croatia2018} \\ \hline
Ecuador & Competence & Competency Framework for Learning & 2023 & Spanish & \cite{Ecuador2023} \\ \hline
Spain & Competence & 1654 Real Decreto, Real Decreto 157/2022, Real Decreto 217/2022 & 2022 & Spanish & \cite{Spain2022, Spain2022b, Spain2022c} \\ \hline
United States & Not applicable & CSTA K–12 Computer Science Standards & 2017 & English & \cite{USA2017} \\ \hline
Finland & ICT Competence & National Core Curriculum for Primary and Lower Secondary (Basic) Education & 2014 & English & \cite{Finland2014} \\ \hline
Greece & Competence & Curriculum for the Basic School ICT Course & 2021 & English & \cite{Greece2021} \\ \hline
Indonesia & ICT Competence & Curriculum in early childhood education, basic and secondary education levels & 2024 & Indonesian & \cite{Indonesia2024} \\ \hline
Ireland & Competence & Computer Science - Curriculum Specification & 2028 & English & \cite{Ireland2028} \\ \hline
Israel & Computer Science & Education 8th Edition & 2019 & English & \cite{Israel2019} \\ \hline
Italy & Not applicable & National Digital School Plan (PNSD) / National guidelines and new scenarios & 2017 & Italian & \cite{Italy2017, Italy2018} \\ \hline
Jordan & Not applicable & Jordan's National Strategy for Human Resource Development 2016-2025 & 2016 & English & \cite{Jordan2016} \\ \hline
Malaysia & Computer Science & Technology in education: a case study on Malaysia & 2023 & English & \cite{UNESCO2023} \\ \hline
Malta & Computer Science & A National Curriculum Framework for All 2012 & 2012 & English & \cite{Malta2012} \\ \hline
Morocco & Computer Science & Ministry of National Education, Official programs and instructions for computer Science education in common cores / Article - Computer Science Program in Moroccan Secondary Schools: Curricula Analysis & 2005 & English & \cite{UNESCO2005, Morocco2019} \\ \hline
Mexico & Competence & Educational Model for Compulsory Education & 2017 & Spanish & \cite{Mexico2017} \\ \hline
Norway & Competence & Digitization Strategy for Basic Education 2017-2021 & 2017 & Norwegian & \cite{Norway2017} \\ \hline
Paraguay & Not applicable & Management of Information and Communication Technologies (ICT) & 2014 & Spanish & \cite{Paraguay2014} \\ \hline
Peru & ICT Competence & National Curriculum for Basic Education & 2016 & Spanish & \cite{Peru2016} \\ \hline
Poland & Competence & Pos. 356 - Law establishing the new curriculum & 2017 & Polish & \cite{Poland2017} \\ \hline
Portugal & Competence & Order No. 8209/2021 - Approves the Essential Learnings of the curriculum/subject component of Mathematics included in the basic general education curriculum matrix, annexes I to III of Decree-Law No. 55/2018, of July 6. & 2021 & Portuguese & \cite{Portugal2021} \\ \hline
Sweden & Computer Science & Curriculum for the compulsory school, preschool class and school-age educare - REVISED 2018 & 2018 & English & \cite{Sweden2018} \\ \hline
Thailand & Computer Science & The Basic Education Core Curriculum & 2008 & English & \cite{Thailand2008} \\ \hline
Taiwan & ICT Competence & Curriculum Guidelines of 12-Year Basic Education & 2014 & English & \cite{Taiwan2014} \\ \hline
Turkey & Competence & BTY Course Curriculum (Primary Education 1st, 2nd, 3rd, and 4th years) & 2018 & Turkish & \cite{Turkey2018} \\ \hline
\end{longtable}

In the basic education curricula of these countries, three types of insertion can be found: a) Computer Science: where subjects or contents related to Computer Science, such as programming, are indicated in the curricula as shown in the table; b) ICT Competence: cases where there is a description of competencies related to the use of information and communication technologies; and c) CT Competence: listed for countries where the explicit term was found in the curricula, for example, within a knowledge area such as Mathematics or even within the area of Computer Science.

In three countries, none of the terms were found in the consulted documents. Italy, Jordan, and Paraguay. Computational competencies are not part of Italy's official and mandatory national curriculum, but there are guidelines issued by the Ministry of Education on digital education that address CT. One of the documents is \textit{Il Piano Nazionale Scuola Digitale (PNSD)} and the other is the \textit{Indicazioni Nazionali e Nuovi Scenari}. In the case of Jordan, national and international initiatives were found: a) Jordan's National Strategy for Human Resource Development 2016-202, which mentions efforts to modernize education, including the use of technology, suggesting the inclusion of digital and CT skills; b) Organizations such as UNESCO and UNICEF have collaborated with Jordan on various educational fronts, including the promotion of STEM education. Since 2008, Paraguay Educa (a private initiative) has been promoting digital education. Currently, the country is working on a plan to introduce CT into the school curriculum. Although there is no specific policy for the implementation of ICT in the curriculum, the country has some initiatives worth highlighting: a) The Virtual Paraguayan Encyclopedia (2007); b) Integration of ICTs in the Paraguayan School System (2008); c) Technological Classrooms in Paraguay (2009); d) Project 1:1 (2010); e) The One Laptop per Child Project in the Areas of Reading and Logical-Mathematical Thinking (2010-2013).

\begin{figure}[h]
\centering
\includegraphics[width=0.9\linewidth]{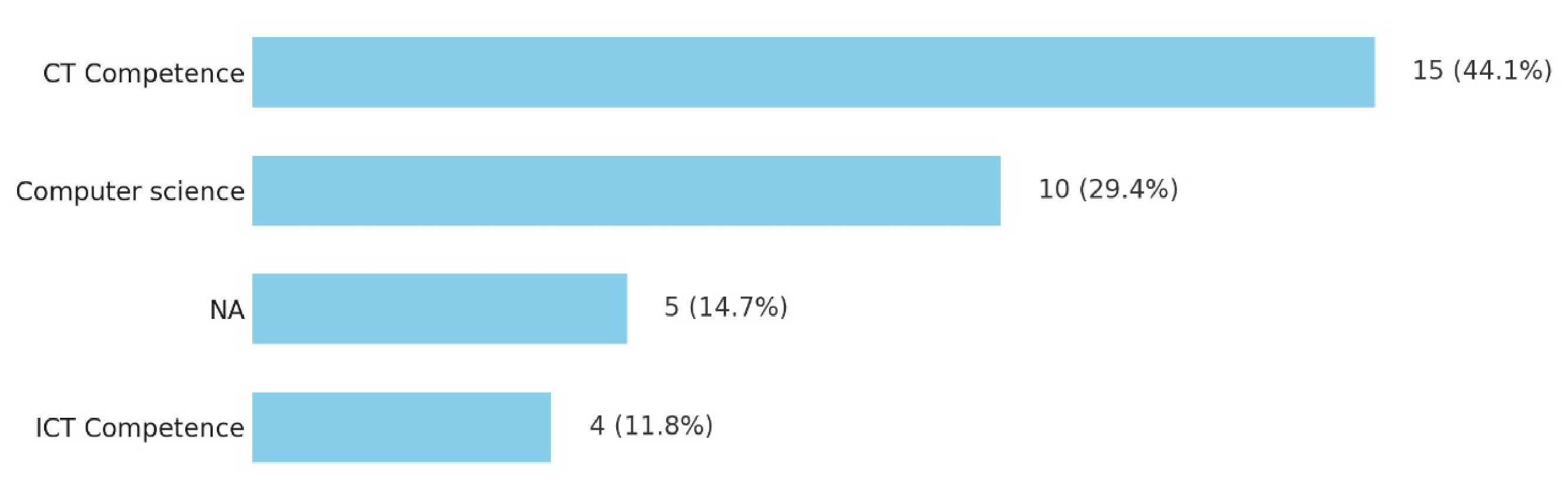}
\Description{Bar chart showing the number of countries by curriculum type. CT Competence has 15 countries (44.1\%), Computer Science has 10 countries (29.4\%), NA has 5 countries (14.7\%), and ICT Competence has 4 countries (11.8\%).}
\caption{Number of countries by curriculum}
\label{fig:countries_curriculum}
\end{figure}

Figure~\ref{fig:countries_curriculum} summarizes the quantitative scenario of countries in each of the categories highlighted under the item "Presence in the school curriculum."

It is therefore observed that only five countries do not have the terms in their curricula, and two of them (the United States and Canada) lack a unified national curriculum. However, in these cases, it was possible to find regional curricula that aim to include CT in their educational programs. The other three countries, Italy, Jordan and Paraguay, have government initiatives aiming to include technologies in educational perspective. From the same figure, it can be inferred that 25 out of the 34 total countries have CT or Computer Science present in some form in their basic education curricula. Lastly, four countries recognize the importance of developing ICT competencies in students.

The inclusion of digital technologies in school curricula is essential for preparing students for the network society described by \citet{bkcastells1999}. In~\cite{arEl-Hamamsy202321}, the authors used advanced statistical analyzes to examine the impact of curriculum reform in Switzerland, which included Computer Science (CS). CT was found to be fundamental to the success of curriculum reform, as it helps equip students with the necessary skills to face the challenges of the 21st century and contributes to a more equitable and inclusive education. The conclusions suggest that the introduction of CS for all students and proper teacher preparation can help reduce performance and perception gaps, promoting equity in the education of CS.

In the current digital age, global interconnection and the fluidity of information profoundly shape social, economic, and cultural life. Most countries recognize this necessity, with 25 of the 34 analyzed incorporating CT or Computer Science into their basic curricula, and others adopting government initiatives to promote technological education. This trend reflects the importance of equipping students with ICT skills, enabling them to navigate and contribute effectively to the contemporary society, which is increasingly based on digital networks and instant communication.

\section{Digital Tools in Teaching Computational Thinking} \label{secTools}

Digital technologies have offered new opportunities for teaching and learning in more interactive and effective ways. They allow access to vast educational resources, personalize the learning experience according to individual student needs, and promote essential 21st century skills such as critical thinking, creativity, and digital literacy. With the continuous advancement of digital technologies, education is becoming more accessible, inclusive, and adaptable, better preparing students for future challenges.

The integration of digital technologies into education is closely linked to the adoption of CT in schools. By using digital tools such as visual programming platforms, educational apps, and virtual learning environments, students have the opportunity to develop CT skills in a practical and engaging way. These technologies facilitate the understanding of complex concepts, allowing students to experiment, create, and solve problems collaboratively and creatively.

In~\cite{arMelanderBowden201958}, the authors focused on children around 10 years old in primary education while exploring problem solving practices in the context of digital activities. Using Scratch software as the central tool, the researchers investigated the multimodal interactions of children as they solved coding problems in a collaborative environment. The study concluded that children learn creative and artistic skills through interaction with digital technologies, positioning them not just as consumers but also as creators and producers of media. Meanwhile, in~\cite{arSahin202386}, it is suggested that students' attitudes towards digital tools, such as programming ones, can influence their motivation and success in the area. Additionally, the use of tools like Blockly and Scratch is considered relevant for integrating technology into the educational environment and developing students' CT skills.

In~\cite{arKucukaydin202349}, the authors used a survey research model to investigate CT skills of 780 third- and fourth-grade students in a large city in Turkey, with the aim of identifying the factors that most influence these skills. The results revealed that the use of technology in the classroom, attitudes towards mathematics, and the educational level of the mother were strongly correlated with the CT skills of the students. 

The adoption of digital technologies in basic education is crucial for developing CT skills that can be applied in various areas of the school curriculum. These skills, which include decomposing complex problems, creating and applying algorithms, and abstracting essential concepts, are fundamental for problem solving and innovation in multiple disciplines. Integrating these practices into the educational environment through interdisciplinary projects and the use of digital technologies not only enriches learning in subjects such as mathematics, science, art, and social studies, but also prepares students for the challenges and opportunities of the digital world. This reflects the importance of preparing students for the network society described by~\citet{bkcastells1999}, where global interconnection and information fluidity are determinants for professional and personal success.

From the literature reviewed in this work, it was possible to identify some competencies and skills related to the implementation of CT in education. According to \cite{arMa202154}, developing cognitive skills such as abstraction, algorithm design, automation, data representation, pattern generalization, debugging and error detection, and non-cognitive skills such as confidence in dealing with complexity, the ability to handle open-ended problems, and the ability to communicate and work in groups to achieve a common goal. Authors like~\cite{arTang202097} define CT skills through various theoretical frameworks, including \citet{inBrennan2012}, which involve computational concepts such as sequences, loops, parallelism, events, conditionals, operators, and data; computational practices such as iteration, debugging, and abstraction; and computational perspectives such as expression, connection, and questioning.

In~\cite{arTan202196}, the authors adopted the STEAM approach (Science, Technology, Engineering, Arts, and Mathematics), combined with the Scratch platform, to develop conceptual games about electricity. The results demonstrated a positive impact of the STEAM approach on improving critical thinking skills, as well as developing five CT skills (algorithmization, cooperativity, creativity, critical thinking and problem solving). There was a significant increase in the average CT scores and the five skills after the intervention. 

\begin{figure}[h]
\centering
\includegraphics[width=\linewidth]{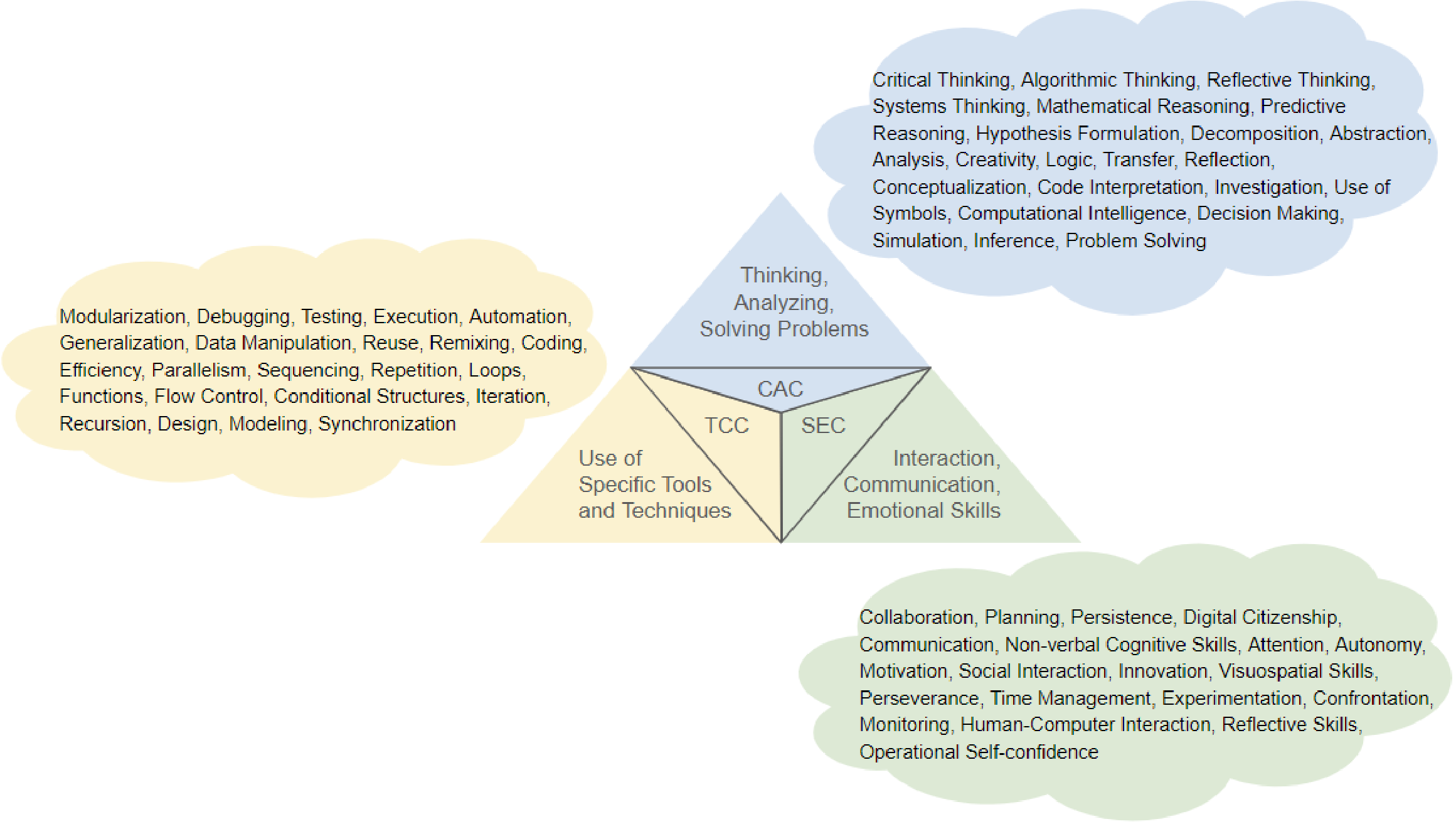}
\Description{Diagram showing the competencies and their associated skills. The competencies are labeled as CAC, TCC, and SEC. The descriptions for these competencies are Thinking, Analyzing, Solving Problems for CAC; Use of Specific Tools and Techniques for TCC; and Interaction, Communication, Emotional Skills for SEC. The skills for each competency are listed in clouds around the diagram. Skills for CAC include Critical Thinking, Algorithmic Thinking, Reflective Thinking, Systems Thinking, Mathematical Reasoning, Predictive Reasoning, Hypothesis Formulation, Decomposition, Abstraction, Analysis, Creativity, Logic, Transfer, Reflection, Conceptualization, Code Interpretation, Investigation, Use of Symbols, Computational Intelligence, Decision Making, Simulation, Inference, and Problem Solving. Skills for TCC include Modularization, Debugging, Testing, Execution, Automation, Generalization, Data Manipulation, Reuse, Remixing, Coding, Efficiency, Parallelism, Sequencing, Repetition, Loops, Functions, Flow Control, Conditional Structures, Iteration, Recursion, Design, Modeling, and Synchronization. Skills for SEC include Collaboration, Planning, Persistence, Digital Citizenship, Communication, Non-verbal Cognitive Skills, Attention, Autonomy, Motivation, Social Interaction, Innovation, Visuospatial Skills, Perseverance, Time Management, Experimentation, Confrontation, Monitoring, Human-Computer Interaction, Reflective Skills, and Operational Self-confidence.}
\caption{The three set of CT competencies, their definitions and skills}
\label{fig:competence}
\end{figure}

Figure~\ref{fig:competence} summarizes the competencies identified in the analyzed literature for this work, indicating which skills were identified in the focused research. According to~\citet{bkPerrenoud2000}, competencies refer to the practical mastery of tasks and situations, which can only be achieved through the development of students' skills and understanding of the content that underpins these domains. For example, for students to master mathematics in their daily tasks, it is necessary to develop their numerical skills, introducing concepts such as number, quantity, and grouping.~\citet{bkPerrenoud2000} differentiates competencies from skills by stating that competencies are practical domains that involve understanding and the purpose of actions, while skills are the specific actions determined by these competencies, such as painting, writing, or playing musical instruments.

It should be noted that the development of such CT skills, as shown in Figure~\ref{fig:competence}, occurred with the support of the following digital tools: @MyClassGame, Alice, Blockly, Blockly Games, Blockly Script, Code.org, Google Colab, Lightbot, Logo, MIT App Inventor, Malt2, Pencil Code, Python Tutor, Scratch, Swift Playgrounds, Tinkercad, and Zoombinis.

Based on the skills identified in the literature, combined with the use of digital tools for the development of CT, it was possible to categorize three main competencies for the application of CT in education: Cognitive and Analytical Competencies (CAC), which reflect a set of skills related to thinking, analysis, and solving complex problems; Technical and Computational Competencies (TCC), which summarize a set of skills related to the use of specific computing and programming tools and techniques; and Social and Emotional Competencies (SEC), which relate to a set of skills involving interaction, communication, and emotional and social aspects.

As explored in this article, the use of digital tools has proven effective in developing CT among students of various ages. These tools provide interactive and practical environments where students can learn programming concepts, logic, and problem solving in an engaging and accessible way. By simulating real-world challenges, the tools allow students to experiment and iterate their solutions, promoting the understanding of algorithms, abstraction, and automation. In addition, many are designed to be intuitive and playful, making learning more attractive and encouraging exploration and creativity.

In~\cite{arMoschella202067}, the study aimed to evaluate the effects of CT training on children's cognitive and creative skills, supported by apps such as LegoWeDo and Scratch. The results indicated improvements in children's spatial and logical skills after the training, suggesting a positive correlation between CT activities and the development of these skills. The authors also emphasized the importance of considering other software, such as Alice and Tynker, in future studies. These findings reinforce the importance of CT as a transversal skill and highlight the potential to integrate this training into various subjects, such as languages, thus expanding interdisciplinary learning opportunities for students. 

In~\cite{arArfe202003}, the objective was to investigate the effects of programming on the planning and response inhibition skills of six-year-old children. Using the Code.org platform, a randomized experiment was carried out with 179 first-grade students in Italy, divided into experimental and control groups. The main contributions of the study include demonstrating that coding activities improved not only children's programming skills but also their executive functions.

In~\cite{arMoreno-Leon202065}, a study was conducted using the Dr. Scratch tool to analyze 250 projects from the Scratch platform, distributed in five main categories: animations, art, games, music, and stories. They proposed a personalized learning path for developing students' CT based on a detailed analysis of Scratch projects. The results suggest that different types of projects can be used to develop various dimensions of CT, providing opportunities to develop programming skills in a variety of educational contexts.

In~\cite{arSanchezCamacho202387}, the goal was to describe educational programs that teach CT in primary education, analyzing the competencies and CT skills addressed. The digital tools covered include Scratch, unplugged activities, Code.org, LEGO® Mindstorms EV3, Alice, LEGO® WeDo 2.0, Beebot, AgentCubes, AgentSheets, Kodu Game Lab, Scratch JR, IchigoJam, Arduino, Turtle Academy, CompThink, EasyLogic 3D, Plastelina interactive Logic Game, Scalable Game Design, Microe Virtual Environment Interactions (VENVI). The results showed that the CT abilities most addressed are problem solving, algorithmic thinking, logical data organization, data representation (abstraction), generalization, and transfer. Open tasks include creating a common framework and the need for more studies exploring the effectiveness of CT educational programs.

\subsection{Classification of Tools}

Some distinctions between the tools are suggested in the literature. In~\cite{arKanika202044}, the authors examined a wide range of articles published from the 1990s to the present, focusing on innovative approaches and tools that impacted student performance. They categorized these tools into five main groups: visual programming, game-based learning, pair and collaborative programming, robotics programming, and assessment systems. In~\cite{arTikva202199}, the research involved a systematic review of the literature, in which the authors examined 47 studies that used or developed tools to teach CT. They classified the tools into three subareas: programming tools and communities, robotics and microcontrollers, and tools specifically developed for CT.

In the present article, we propose a classification based on previous work augmented by observations on the current state-of-the-art. Similarly to~\cite{arKanika202044}, we use the category of Visual Programming, but in the literature there is a considerable variety of Textual Programming tools validated for teaching CT, although there is evidence that Visual Programming is more effective for teaching children and people who are not in the technology field~\cite{arHijon-Neira202335,arHou202036}. Therefore, we suggest subdividing the Programming category into Textual and Visual modalities. We do not consider the category of collaborative programming, as in~\cite{arKanika202044, arTikva202199}, since most of the analyzed tools have some feature that enables synchronous or asynchronous collaboration between users. We also consider the category of Electronic Games as in~\cite{arKanika202044} and Robotics as in both~\cite{arKanika202044, arTikva202199}. Finally, we add the categories of Modeling and Simulation and Unplugged, which are types of tools widely available in the literature for the CT context.

\begin{figure}[h]
\centering
\includegraphics[width=0.85\linewidth]{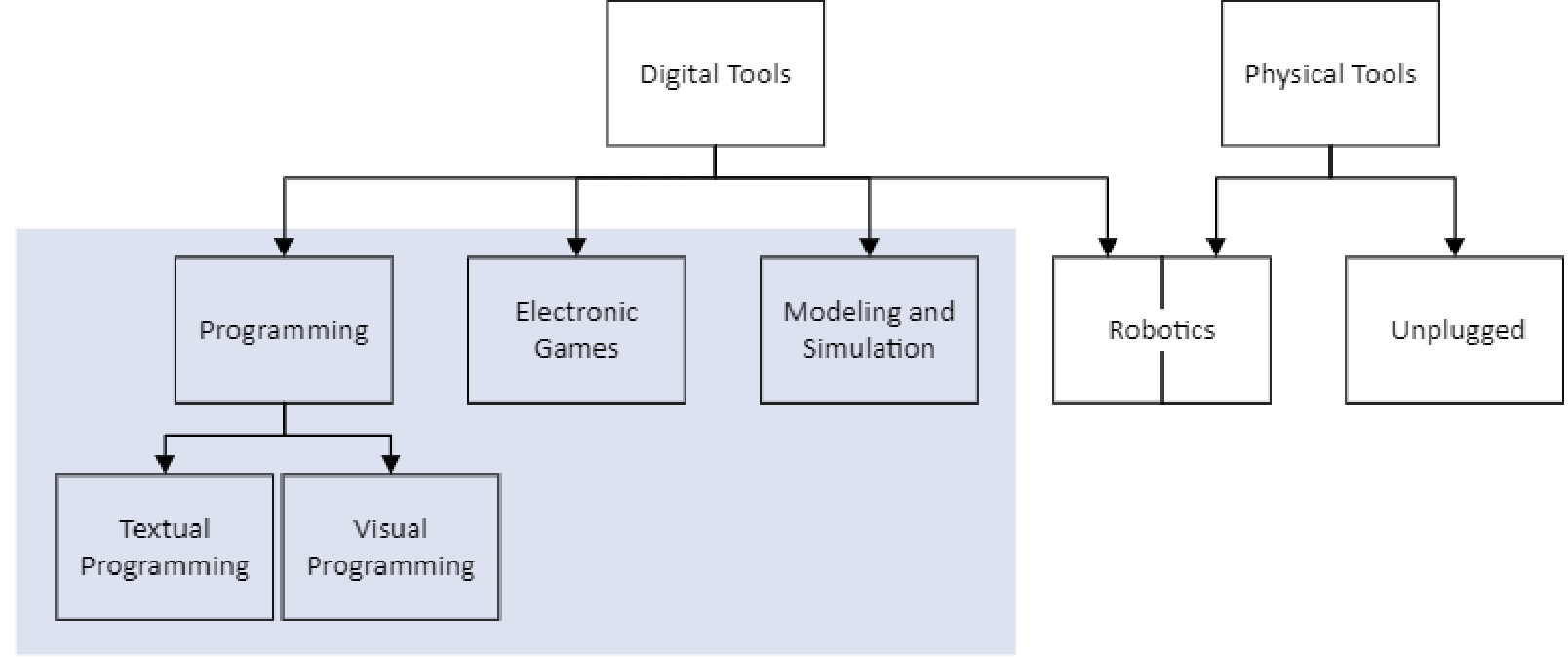}
\Description{Diagram showing types of tools commonly found in the literature in the context of CT, highlighting those considered in the present research. The diagram is divided into Digital Tools and Physical Tools. Digital Tools include Programming (with subcategories Textual Programming and Visual Programming), Electronic Games, Modeling and Simulation, and Robotics. Physical Tools include Robotics and Unplugged.}
\caption{Types of tools commonly found in the literature in the context of CT, highlighting those considered in the present research}
\label{fig:classes}
\end{figure}

Figure~\ref{fig:classes} shows the proposed categories for identifying CT tools and situates the focus of the present investigation on the highlighted blocks, which include tools in the categories of programming, electronic games and modeling and simulation.

It is worth noting that some tools can be considered both physical and digital, such as programmable robotics kits. This justifies the subdivision of the Robotics category block in Figure~\ref{fig:classes}.

\subsection{Digital Tools Used in Computational Thinking}

In this section, we examine the relevant characteristics of each tool found in the literature to teach CT in the categories considered.

\subsubsection{Python Tutor}

It can be described as an educational tool for Python. Its main features are: step-by-step visualization of code execution; interactive debugging; support for different versions of Python.

This tool has the potential to deepen the understanding of Python concepts, teach error identification and correction, and promote autonomy and self-learning. The literature reports positive outcomes from its use at the K-12 level for teaching programming, as graphical visualization of code execution can mitigate potential abstraction difficulties that hinder learning. In addition, it helps promote problem solving skills, algorithmic thinking, and logical reasoning in this age group~\cite{arMladenovic202159}.

\subsubsection{Google Colab}

It is a collaborative development environment that primarily supports Python but can also support other languages such as R, Julia, Swift, etc. The main features of this tool are: editing and running Python notebooks in the cloud; storing and sharing notebooks; collaborating on notebooks with others; access to hardware accelerators.

In education, it can be applied to promote collaborative learning, develop general programming projects, or even specific projects such as data science and machine learning, as well as those involving high-performance processing with large volumes of data. Although its use requires an Internet connection, there is no need for software installation and configuration on the device on which it will be used, as it is a web application. This facilitates interaction between the student and the tool in environments outside the educational institution or even remote higher education, continually enhancing CT skills such as decomposition, algorithmic thinking, and pattern recognition~\cite{arVidal-Silva2022109}.

\subsubsection{Blockly}

A versatile tool that can be used as a visual programming editor or as a library that can be integrated into other tools. It generates code for various languages including JavaScript, Python, PHP, Lua, Dart, and others, depending on the configuration. Its main features are: the ability to drag and drop blocks to create code; support for multiple programming languages; and use in creating games, animations, web applications, etc.

In an educational context, it can be used to introduce basic programming concepts in a playful way, develop logic, CT, and creativity, and promote project-based learning. The literature reports significant improvements in CT skills of K-12 students both technically, such as understanding concepts such as algorithms, loops, and debugging, and in the promotion of social and emotional skills such as collaboration, persistence, and creativity~\cite{arTran2019100}. However, in~\cite{arUnal2021104}, authors found that the tool does not have a substantial impact on ``programming anxiety'' in high school students. In the university context, the integration of Blockly with Jupyter Notebook was explored to understand student interaction and perception about the design of a block-based programming tool, concluding that the design of the tool significantly influences learning experiences~\cite{arTawfik202498}.

Blockly is among the six most discussed digital tools in the literature, considering the categories addressed in this research for the CT context between 2019 and 2024. Consequently, various CT skills that can be positively influenced by its use were analyzed, such as: modularization, logic, debugging, automation, generalization, iterative reasoning, critical thinking, algorithmic thinking, decomposition, abstraction, collaboration, persistence, hypothesis formulation, pattern recognition, problem-solving, and planning~\cite{arKuz202250, arMorze202266, arVanicek2022108, arTran2019100, arTawfik202498, arUnal2021104, arTikva202199}.

\subsubsection{BlocklyScript}

This is an extension of Blockly for JavaScript. Its main features are: generating JavaScript code from visual blocks; executing JavaScript code in the browser; debugging JavaScript code.

In education, BlocklyScript can be used to teach JavaScript programming visually, enabling students to develop websites, games, and other interactive applications. It is also useful for developing debugging and code testing skills, and for promoting autonomy and self-learning. K-12 students have reported a positive experience using this tool to enhance CT skills such as algorithmic thinking, modularization, problem solving, generalization, abstraction, and pattern recognition~\cite{arKarakasis202045}.

\subsubsection{Blockly Games}

It can be defined as a web platform that offers a set of educational games. This tool allows users to: play educational games to learn programming; solve logic and CT challenges.

Blockly Games can motivate programming learning in a playful manner, increasing engagement; develop logic, reasoning, and problem solving skills; and introduce basic programming concepts. In the literature, it was used with 10th-grade students and showed positive results in mitigating apathy and demotivation in the classroom~\cite{arDiaz-Lauzurica201919}.

\subsubsection{Alice}

It is a 3D programming environment that uses a block-based visual language, generating code in Java. Its main features include: drag-and-drop 3D objects to create animations; program behaviors and interactions with scripts; create games, interactive stories, etc.

In an educational context, it can be used to teach programming and 3D design concepts in an integrated manner, develop creativity and storytelling skills, and promote interdisciplinary learning. The effectiveness of Alice in improving the performance of K-12 students in programming has been reported in~\cite{arHu202139}. This tool can positively influence participation, reflection, collaboration, and problem solving, which benefit CT education, especially for the K-12 age group~\cite{arBasogain201806, arDurak2020117}.

Like Blockly, Alice was one of the six tools most discussed in the literature, considering the categories addressed in this research for the CT context between 2019 and 2024, in the context of K-12, undergraduate (UG), and teacher training. Its potential to improve CT skills has been discussed, including modularization, debugging, automation, generalization, iterative reasoning, creativity, data manipulation, critical thinking, logic, decomposition, abstraction, collaboration, problem solving, hypothesis formulation, operational self-confidence, algorithmic thinking, transfer, and planning \cite{arDurak2020117, arBasogain201806, arHu202139, arMorze202266, arTikva202199, arSanchezCamacho202387, arSun202295}.

\subsubsection{MIT App Inventor}

It is a programming environment that uses a visual block language based on Blockly, translating into code to develop Android applications. Its main functionality is the ability to create Android apps by dragging and dropping blocks.

MIT App Inventor can be used to teach the development and design skills of mobile apps. In the literature, it is the second most analyzed tool between 2019 and 2024, considering the categories addressed in this research, in the contexts of K-12, UG, and teacher training. In summary, studies that cited this tool highlight its potential in enhancing CT skills such as modularization, debugging, testing, uncertainty tolerance, generalization, iterative reasoning, creativity, reuse, remixing, critical thinking, logic, decomposition, abstraction, attention, autonomy, collaboration, motivation, persistence, communication, algorithmic thinking, problem solving, and social interaction \cite{arHu202139, arTikva202199, arSantos202389, arCalandra202111, arHsu202137, arRochadiani202381, arAttard202005, arHsu202238, arKong202347, arPerez-Jorge202277, arRich202280}.

\subsubsection{Scratch}

It is a visual programming environment. Among its features are: drag-and-drop blocks to create code; facilitate the creation of games and animations with blocks for control, movement, appearance, sound, and sensing; hardware integration; online project sharing with the community. This is the most used tool in the context of CT.

In an educational context, it can be used to introduce programming in a playful manner, develop logic, CT, and creativity, promote collaborative learning, and support project-based learning. In~\cite{arKanika202044}, the authors examined a wide range of articles published from the 1990s to the present and revealed that Scratch was the most researched and cited tool, with more than 1,000 citations, highlighting its broad adoption and recognition in the academic community. This is due to its effectiveness and versatility in various teaching practices that enhance multiple CT skills~\cite{arIbrohim202340}, and reports of its use in various age groups. During the past 5 years, it has been predominantly researched in the K-12 context \cite{arSun202295, arFAGERLUND202026, arPiedade202279}, with some occurrences in undergraduate studies \cite{arGamito202231, arHou202036, arSilva201990}, and teacher training, as in~\cite{arGABRIELE201930}, and one occurrence of using ScratchJr in the preschool context~\cite{arKourti202348}.

There are reports of significant benefits in creativity, cooperative and critical thinking skills associated with the use of Scratch~\cite{arJiang202141, arPerez-Marin202078}. In~\cite{arHandayani202334, arSilva202192, arWong2024112}, the tool is mentioned for its use in collaborative and problem-solving approaches. In \cite{arMa202154, arWei2021111}, the tool was able to stimulate critical thinking, also addressed in~\cite{arPark202274}, and student self-efficacy. In~\cite{arBroza202309}, the use of the tool brought benefits to the abstraction and decomposition skills. Given the numerous CT skills enhanced with the help of Scratch reported in the literature over the years, it is simpler to approach it as a tool capable of positively influencing social and emotional competencies, technical and computational competencies, and cognitive and analytical competencies.

\subsubsection{LightBot}

It is an educational programming game. Specifically, the tool offers puzzles that can be solved using programming commands.

LightBot can be used to develop logic and problem solving skills, introduce programming gradually (with different levels of difficulty), and support autonomous and individualized learning. In~\cite{arYallihep2020116}, it was used in an Information Technology and Software course for K-12 students. The use of LightBot resulted in a significant increase in student performance in programming concepts. The tool was also validated with undergraduate students, showing good gains in improving CT skills such as decomposition, generalization, and problem solving \cite{arLee202052, arUrquizo2021106}. In~\cite{arSantos202388}, the successful use of the tool in early childhood education is reported, with advances in problem solving, critical thinking, and creativity.

LightBot was the fourth most cited digital tool in the literature, considering the categories addressed in this research for the CT context between 2019 and 2024. Thus, its potential to improve CT skills has been mentioned, including reuse, modularization, sequencing, critical thinking, algorithmic thinking, debugging, repetition, problem solving, logic, decomposition, generalization and abstraction \cite{arKuz202250, arYallihep2020116, arLee202052, arUrquizo2021106, arSantos202388, arMagnoDeJesus202055, arBedar202001, arErsozlu202323, arPapadakis202172}.

\subsubsection{Tinkercad}

It is a 3D modeling and simulation tool that also supports code editing in visual language and C/C++ for Arduino. Its features include: drag-and-drop blocks to create 3D shapes; edit and modify objects with precise tools; simulate the operation of mechanisms and assemblies; share projects online with the community; prototype and 3D print created projects; edit code based on the Arduino IDE; use a block-based visual programming interface that generates the corresponding Arduino code.

In the educational context, it enables the learning of basic design and 3D modeling concepts, the development of spatial and problem solving skills, preparation for engineering projects, prototyping and simulation of electronic circuits, and learning robotic concepts in conjunction with Arduino. Scientific productions related to the use of this tool are generally associated with Arduino. However, in~\cite{arEryilmaz202124}, the focus of the research was on teaching programming and developing CT skills for students from 5th to 8th grade using Tinkercad, which showed positive results that impacted the learning process. Studies point to Tinkercad's ability to aid in the enhancement of CT skills such as creativity, critical thinking, algorithmic thinking, collaboration, problem-solving, logic, planning, generalization, automation, decomposition, abstraction, and hypothesis formulation \cite{arMorze202266, arSantos202389, arEryilmaz202124}.

\subsubsection{Zoombinis}

This tool falls into the category of Electronic Games. It offers puzzles and logical challenges to practice problem solving.

In the educational context, it can be used to help develop logic and problem-solving skills, facilitate learning through games and interactive challenges, foster critical and creative thinking, and introduce mathematical logic. There are reports of its use in contexts ranging from the 3rd to the 8th grade, with results showing improvements in student CT practices \cite{arAsbell-Clarke202104, arRowe202184}.

\subsubsection{Malt2}

It can be defined as a programming environment and a tool for creating and manipulating 3D graphic models. It allows programming in both visual and textual formats, creating 3D models, and exploring applied mathematical concepts.

In education, it can be used for introducing programming, developing logic, CT, and creativity, learning mathematics, and developing robotics projects. The literature contains several articles demonstrating the effectiveness of the tool in teaching CT to students from 8th to 10th grade through the development of games, 3D modeling~\cite{arGrizioti202133}, and the creation of animated 3D drawings~\cite{arKYNIGOS201851}. It enhances algorithmic thinking, analysis, decomposition, abstraction, and pattern recognition.

\subsubsection{@MyClassGame}

This tool can be described as an educational gamification platform that has the functionality to create customized educational games.

In the educational context, it can make learning more dynamic and interactive, promote formative assessment, and reinforce curricular content in a playful manner. @MyClassGame also has the potential to help develop skills such as decomposition, abstraction, algorithmic thinking, and analysis, and to be a component of the gamification of the learning process, as analyzed in~\cite{arOlmo-Munoz202369} with students in the second grade.

\subsubsection{Code.org}

It can be defined as a programming learning platform. Its features include the availability of courses, lesson plans, support materials, assessment tools, challenges, and interactive activities on different programming languages, both visual (based on Blockly) and textual (JavaScript, Python, HTML/CSS, etc.).

Code.org can be used to introduce programming comprehensively, develop essential computational skills, and facilitate learning through interactive and gamified activities. The literature reports that using Code.org positively impacted non-verbal cognitive skills of PK and KG children~\cite{arCiftci202016}. In~\cite{arChen202314}, the authors mention that a programming course conducted with Code.org had a positive impact on the development of CT and student motivation to learn programming, especially in the self-directed study modality for 3rd to 4th-grade children. In~\cite{arArfe202003}, the tool was applied to PK and 1st-grade children. The research involved the application of standardized tasks to assess children's cognitive skills before and after their exposure to coding activities on Code.org. 

Code.org was the third most cited digital tool in the literature between 2019 and 2024 in the context of CT, considering the categories addressed in this investigation (Figure \ref{fig:countries_curriculum}). It has been applied from PK to higher education and teacher training, aiding in the development of CT skills such as: modularization, reflection, logic, mathematical reasoning, critical thinking, algorithmic thinking, pattern recognition, debugging, reuse, resilience, automation, analysis, generalization, creativity, data manipulation, decomposition, abstraction, data representation, confrontation, collaboration, digital citizenship, visual memory, persistence, communication, non-verbal cognitive skills, operational self-confidence, transfer, problem-solving, and spatial intelligence \cite{arTran2019100, arSanchezCamacho202387, arErsozlu202323, arCiftci202016, arChen202314, arArfe202003, arKale202142, arKale202343, arCutumisu201917, arMcCormick202256}.

\subsubsection{Logo}

It is an educational programming language that allows programming graphics, drawings, and animations using simple commands.

Logo supports interactive learning of mathematical and geometric concepts, developing skills in monitoring, reflection, conceptualization, algorithmic thinking, debugging, coding, testing, analysis, execution, decomposition, introducing textual programming and promoting interdisciplinary learning (areas such as mathematics, geometry, programming and art) \cite{arTrinchero2019101, arValentine2018107}. In~\cite{arTrinchero2019101}, the authors used this language to develop Edulogo, an online game, as the main tool to introduce CT to children from the 1st to 5th grade. The results were positive, demonstrating the potential of Logo for both direct student use and the development of other tools that can also be utilized in the CT context.

\subsubsection{Swift Playgrounds}

This tool is a programming learning application in the Swift language for creating applications for Apple devices. Its features include the ability to drag and drop blocks to create Swift code, support for different types of applications and games, and the option to share projects online with the community.

In the educational context, it facilitates the playful introduction to iOS app programming, the development of logic, CT, and creativity, learning basic concepts of interface design and user experience, and preparing for the development of basic iOS applications. This tool showed good results in~\cite{arCheng202115}, where it was evaluated by fifth-grade teachers and students, and demonstrated benefits in algorithmic thinking, inference, problem solving, logic, decomposition, and pattern recognition skills.

\subsubsection{Pencil Code}

It is a development environment for the web using CoffeeScript (based on JavaScript). Its main features are: developing web applications with instant visualization of code results; support for educational projects; tutorials.

In the educational context, it facilitates the introduction to web development and the creation of interactive digital art projects. In~\cite{arDeng202018}, it is reported that the use of the tool improved both CT knowledge and students' self-confidence and enjoyment in learning programming for 9th to 11th graders, aiding in the enhancement of skills such as reuse, modularization, algorithmic thinking, debugging, testing, remixing, and abstraction.

\subsubsection{Overview of Tools}

Table~\ref{tab:tools_overview} consolidates the information about each digital tool addressed in the present research. 

\begin{longtable}{|p{2.4cm}|p{2.2cm}|p{1.9cm}|p{1.7cm}|p{1.7cm}|p{3cm}|}
\caption{Overview of digital tools used in CT education.}
\label{tab:tools_overview} \\
\hline
\textbf{Tool} & \textbf{Category} & \textbf{Platform} & \textbf{License} & \textbf{Age Range} & \textbf{References} \\ \hline
\endfirsthead
\caption[]{(Continued)} \\
\hline
\textbf{Tool} & \textbf{Category} & \textbf{Platform} & \textbf{License} & \textbf{Age Range} & \textbf{References} \\ \hline
\endhead
\hline
\endfoot
Python Tutor & Text & Web & MIT & K-12 & \cite{arMladenovic202159} \\ \hline
Google Colab & Text & Web & Free (with paid options) & UG & \cite{arVidal-Silva2022109} \\ \hline
Blockly & Text/Visual & Web & Apache 2.0 & K-12, UG, Teachers & \cite{arKuz202250, arMorze202266, arVanicek2022108, arTran2019100, arTawfik202498, arUnal2021104, arTikva202199} \\ \hline
BlocklyScript & Text/Visual & Web & Apache 2.0 & K-12 & \cite{arKarakasis202045} \\ \hline
Blockly Games & Visual/Electronic Games & Web & Apache 2.0 & 10 & \cite{arDiaz-Lauzurica201919} \\ \hline
Alice & Text/Visual & Windows and MAC & Creative Commons Attribution-Non Commercial 3.0 & K-12, UG, Teachers & \cite{arDurak2020117, arBasogain201806, arHu202139, arMorze202266, arTikva202199, arSanchezCamacho202387, arSun202295} \\ \hline
MIT App Inventor & Visual & Web; Windows, MAC, and Linux; Android and IOS & Apache 2.0 & K-12, UG, Teachers & \cite{arHu202139, arTikva202199, arSantos202389, arCalandra202111, arHsu202137, arRochadiani202381, arAttard202005, arHsu202238, arKong202347, arPerez-Jorge202277, arRich202280} \\ \hline
Scratch & Visual & Web; Windows, MAC, and Linux; Android and IOS (ScratchJr) & MIT & PK-KG, K-12, UG, Teachers & \cite{arDurak2020117, arBasogain201806, arHu202139, arKuz202250, arMorze202266, arTikva202199, arSanchezCamacho202387, arSun202295, arSantos202389, arKong202347, arPerez-Jorge202277, arRich202280, arIbrohim202340, arFAGERLUND202026, arPiedade202279, arGamito202231, arHou202036, arSilva201990, arGABRIELE201930, arKourti202348, arJiang202141, arPerez-Marin202078, arHandayani202334, arSilva202192, arWong2024112, arMa202154, arWei2021111, arPark202274, arBroza202309, arErsozlu202323, arPapadakis202172, arMontiel202164, arFidai202029, arXing2021115, arELOY202122, arZampieri2020118, arAltanis201902, arChekour202313, arZhang2019120, arSilva202291, arDuo-Terron202320, arXie2023114, arRodriguez-Martinez202083, arMukasheva202168, arFagerlund202127, arMoschella202067, arWang2022110, arChai202112, arPapavlasopoulou201973, arPellas202376, arSaez-Lopez202085, arLiu202353, arEstevez201925, arRodriguez-Benito202082, arGokce202332, arParsazadeh202175, arBender202207, arMolina-Ayuso202260, arPanskyi201971, arMelanderBowden201958, arMonjelat201962, arSjoberg202093, arHijon-Neira202335, arTucker-Raymond2021103, arMolina-Ayuso202361, arMonjelat202063, arMoreno-Leon202065, arKilhamn202246, arBroley202308, arWu2023113, arTan202196, arOzcan202170} \\ \hline
LightBot & Electronic Games & Web; Android and IOS & MIT & PK, K-12, UG, Teachers & \cite{arKuz202250, arYallihep2020116, arLee202052, arUrquizo2021106, arSantos202388, arMagnoDeJesus202055, arBedar202001, arErsozlu202323, arPapadakis202172} \\ \hline
Tinkercad & Modeling and Simulation & Web; Android and IOS & Free for personal, educational, or non-profit commercial use & 5-8, Teachers & \cite{arMorze202266, arSantos202389, arEryilmaz202124} \\ \hline
Zoombinis & Electronic Games & Web; Windows and MAC; Android and IOS & MIT & 3-8 & \cite{arAsbell-Clarke202104, arRowe202184} \\ \hline
Malt2 & Text/Visual & Web; Windows and MAC & MIT & 8-10 & \cite{arGrizioti202133, arKYNIGOS201851} \\ \hline
@MyClassGame & Electronic Games & Web & Free for personal and educational use (Paid plan for schools) & 2 & \cite{arOlmo-Munoz202369} \\ \hline
Code.org & Text/Visual & Web & Creative Commons Attribution 4.0 International & PK-KG, K-9, Teachers & \cite{arTran2019100, arSanchezCamacho202387, arErsozlu202323, arCiftci202016, arChen202314, arArfe202003, arKale202142, arKale202343, arCutumisu201917, arMcCormick202256} \\ \hline
Logo & Text/Visual & Web; Windows, MAC, and Linux & Logo Online: GPLv3 & 1-5, Teachers & \cite{arTrinchero2019101, arValentine2018107} \\ \hline
Swift Playgrounds & Visual & IOS and MAC & Free (some paid features) & 5 & \cite{arCheng202115} \\ \hline
Pencil Code & Text/Visual & Web; Windows, MAC, and Linux & MIT & 9-11 & \cite{arDeng202018} \\ \hline
\end{longtable}

Through Table~\ref{tab:tools_overview}, it can be seen that most tools fall into a hybrid category that allows both visual and textual programming. In addition, most of them have web applications. This is advantageous as they do not need to be installed or configured locally on a device but do require an internet connection. Furthermore, most of the tools are free, at least in the educational context. Finally, all the tools are suitable for the K-12 age range, fully or partially. Some have been researched in higher education and teacher training. Few have been used in PK and KG, namely Scratch, LightBot, and Code.org.

Table~\ref{tab:knowledge_areas} shows the areas of knowledge associated with the use of digital tools in the literature over the past 5 years, considering only the categories addressed. It should be noted that programming and CT are the most common areas for almost all tools. It is interesting to see the variety of knowledge areas that can be integrated into CT teaching. It is also possible to find subjects such as languages, arts, sciences, and even geography. The most widely used tools also have a wide spectrum of knowledge areas.

\begin{table}[h]
\centering
\caption{Knowledge areas in which each tool has been used in CT in the last 5 years}
\label{tab:knowledge_areas}
\begin{tabular}{|p{3cm}|p{11cm}|}
\hline
\textbf{Tool} & \textbf{Knowledge Area} \\ \hline
Python Tutor & Programming \\ \hline
Google Colab & Programming \\ \hline
Blockly & Programming, Languages, Mathematics, Natural Sciences, Computing, Computational Thinking \\ \hline
BlocklyScript & Programming \\ \hline
Blockly Games & Robotics \\ \hline
Alice & Programming, Languages, Mathematics, Natural Sciences, Computing, Computational Thinking, Information Technology and Software \\ \hline
MIT App Inventor & Programming, Mathematics, Artificial Intelligence, Computational Thinking, Psychology, Computing, Engineering \\ \hline
Scratch & Economics, Chinese, Art, Portuguese Language, Computational Thinking, Social Sciences, Computing, Music, Geography, Geometry, Mathematics, Artificial Intelligence, Literature, Natural Sciences, STEM, Education, Engineering, Programming, English Language, Arts, Psychology, Information Technology and Software, STEAM \\ \hline
LightBot & Geography, STEM, Programming, Systems Engineering, Algorithms, Computational Thinking, Information Technology and Software \\ \hline
Tinkercad & Programming, Languages, Mathematics, Natural Sciences, Computing \\ \hline
Zoombinis & STEM, Computational Thinking \\ \hline
Malt2 & Mathematics, Physics, Computing \\ \hline
@MyClassGame & Mathematics \\ \hline
Code.org & STEM, Programming, Natural Sciences, Mathematics, Languages, Computing, Engineering, Computational Thinking \\ \hline
Logo & Mathematics \\ \hline
Swift Playgrounds & Computational Thinking \\ \hline
Pencil Code & Programming \\ \hline
\end{tabular}
\end{table}

\begin{figure}[h]
\centering
\includegraphics[width=0.8\linewidth]{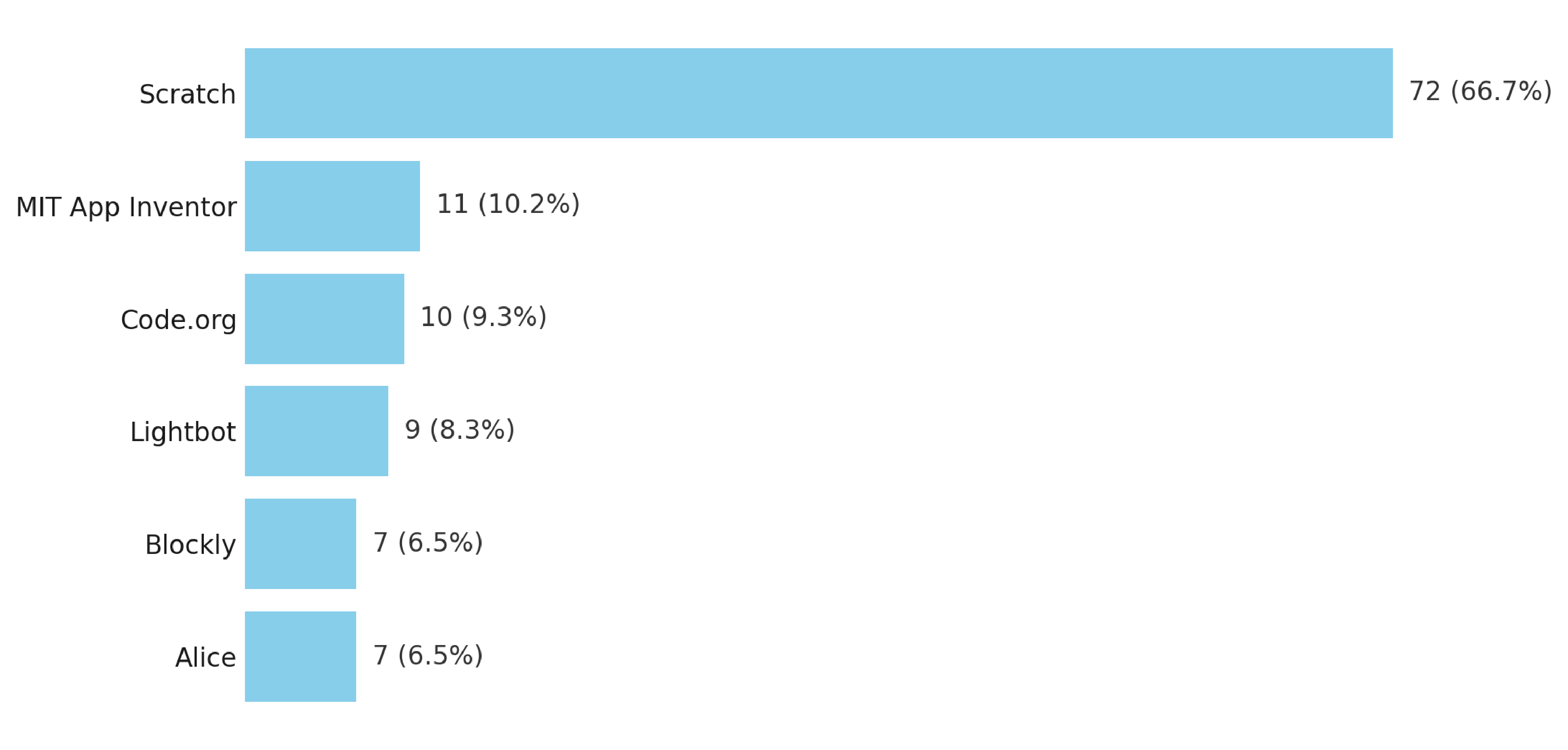}
\Description{Bar chart showing the percentage of use of each tool with at least 5 use cases in the articles analyzed in this research. Scratch has the highest usage with 72 cases (66.7\%), followed by MIT App Inventor with 11 cases (10.2\%), Code.org with 10 cases (9.3\%), Lightbot with 9 cases (8.3\%), Blockly with 7 cases (6.5\%), and Alice with 7 cases (6.5\%).}
\caption{Percentage of use of each tool (with at least 5 use cases) in the articles analyzed in this research}
\label{fig:tool_usage}
\end{figure}

 Figure~\ref{fig:tool_usage} shows that Scratch has an absolute predominance in the analyzed works, being present in 66.7\% of the research (72 of the 108 articles involving the use of digital tools in the context of the present research). In~\cite{arSanchezCamacho202387}, he authors attribute this popularity to several qualities of the tool, including its accessibility, ease of use, and an active user community that facilitates interactive and progressive learning in block programming. The same arguments are reinforced in~\cite{arMontiel202164}, which also highlights the significant role of Scratch in the development of CT skills, providing an engaging and accessible learning environment for students.

\begin{figure}[h]
\centering
\includegraphics[width=0.5\linewidth]{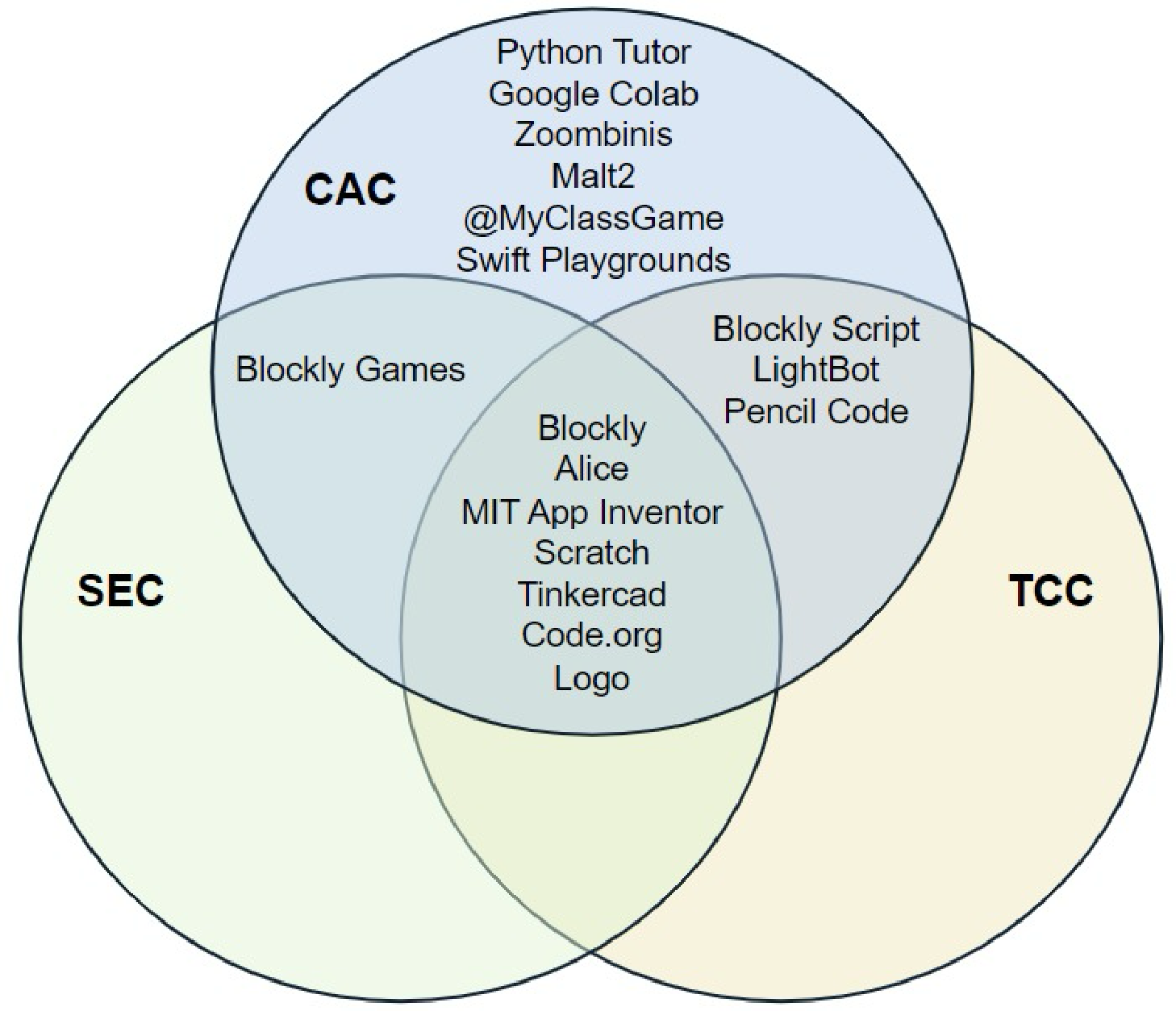}
\Description{Venn Diagram showing competencies addressed by each tool in experiments and literature analyses over the last 5 years. The tools are grouped into three categories: CAC, TCC, and SEC. Various tools are listed within each category, such as Python Tutor, Google Colab, Zoombinis, Malt2, @MyClassGame, and Swift Playgrounds for CAC; BlocklyScript, LightBot, and Pencil Code for TCC and CAC; and Blockly Games for both CAC and SEC. Another group includes Blockly, Alice, MIT App Inventor, Scratch, Tinkercad, Code.org, and Logo for TCC, CAC and SEC.}
\caption{Competencies addressed by each tool in experiments and literature analyses over the last 5 years}
\label{fig:tools_competencies}
\end{figure}

Figure~\ref{fig:tools_competencies} shows the competencies that can be enhanced with the help of digital tools. It is evident that in addition to the six most widely used tools, Logo and Tinkercad can also positively influence all three evaluated competencies (CAC, TCC, and SEC). Furthermore, it is important to emphasize that the competencies are much more associated with the activities that students engage in with the help of the tools, rather than the tools themselves. Thus, a tool like Google Colab, for example, which is marked only in CAC, could still be capable of exercising the other competencies. In other words, there have simply not been, in the literature of the last 5 years, analyses of studies that proposed activities with Google Colab, in the context of CT, for the enhancement of TCC and SEC. This also applies to the other tools.

\section{Research Opportunities} \label{secOpportunities}

The objective of this section is to elucidate challenges and gaps in the state-of-the-art, aiming to stimulate scientific productions capable of mitigating them and bringing advances in the teaching of CT using digital tools.

\subsection{Infrastructure}

Infrastructure issues are relevant topics in the context of the use of digital tools in education \cite{arDurak2020117, arVidal-Silva2022109, arEryilmaz202124, arCheng202115, arStupuriene202494}. Some tools require an internet connection, and all of them run on devices such as computers, smartphones, and tablets, which are not always available in sufficient quantity or in adequate condition. In~\cite{arStupuriene202494} it is suggested that sufficient resources should be allocated to ensure that schools have the necessary digital tools and infrastructure to support effective technology integration. However, this is not a simple goal in all countries or regions. This problem is complex and can be intensified or mitigated by economic and political factors. The solution is context-dependent but can be facilitated if there are well-defined guidelines for teaching CT, with clear definitions of needs and objectives.

In \cite{arVidal-Silva2022109}, challenges for using Google Colab were pointed out, precisely related to infrastructure issues, as there is a need for internet access, and the learning curve for its use. When these obstacles are overcome, the results tend to be positive, according to the research, especially in the development of basic and advanced computational skills. Despite this, the potential of this tool in teaching CT is still underexplored in the literature.

In~\cite{arDurak2020117}, the authors pointed out that infrastructure outside the school environment is a challenge for practicing the concepts learned in class at home. This is a factor that can also negatively impact student engagement.

\subsection{Difficulties in Student Engagement and Use of Digital Tools}

Development and use projects for educational games can motivate students to develop CT skills~\cite{arKarakasis202045}. Converging with this idea, in~\cite{arDiaz-Lauzurica201919}, the authors, through action research, observed that a teaching methodology proposed using Blockly Games was effective in increasing students' motivation and engagement, as well as developing their CT and programming skills. This was also observed in \cite{arPellas202376, arLiu202353} through the use of Scratch. According to the authors of~\cite{arPanskyi201971}, the use of games can advance CT knowledge even outside class hours and environments.

This positive influence of games on CT education was the main focus of~\cite{arOlmo-Munoz202369}, in which the authors used the @MyClassGame tool with 82 second-year students. The objective was to investigate the impact of superficial and deep gamification techniques on students' CT skills and intrinsic and extrinsic motivation. The positive impact was proven, especially when deep gamification is considered. However, there are few articles in the literature that explore the use of this tool.

In~\cite{arYallihep2020116, arLee202052, arUrquizo2021106, arSantos202388, arBedar202001}, the authors also explored the use of games using the LightBot tool. All of these studies demonstrate the effectiveness of using the tool to improve problem solving capacity, critical thinking, and creativity of students of different age groups, including undergraduates. However, it can be challenging to adapt this tool to effective pedagogical practices~\cite{arBedar202001}, \cite{arUrquizo2021106}. According to the authors of~\cite{arSantos202388}, the game is repetitive after a certain time, lacks customization options or adaptation of difficulty levels to meet individual learning needs. Clearly, despite its effectiveness, there is a gap in the literature on more adequate exploration of the tool, as well as on consideration of its limitations according to the target discipline and age group.

In~\cite{arMagnoDeJesus202055}, a framework is proposed for teaching CT through Problem-Based Learning and Digital Games strategies, such as LightBot. An interesting point of this research is the identification of factors that, if promoted, contribute to student engagement. These factors are collaboration, individual responsibility, shared leadership, direct social skills, and intrinsic motivation.

Another tool used in the context of games to teach CT is Zoombinis. In~\cite{arAsbell-Clarke202104, arRowe202184}, the authors observed improvements in students' CT practices after their use. In~\cite{arRowe202184}, a tool is presented to use the data automatically collected during the execution of the game to assess the implicit CT knowledge acquired by the students. However, in~\cite{arAsbell-Clarke202104}, the authors highlight the need for further investigation of this knowledge. Although the use of Zoombinis alone has shown improvements in students' CT practices, the contributions to educational practices aimed at facilitating the transfer of knowledge and skills acquired from a learning context to other contexts or situations have not been clearly established.

Malt2 is also a tool with proven effectiveness in promoting CT knowledge~\cite{arKYNIGOS201851}. In~\cite{arGrizioti202133}, the authors use this tool and highlight the challenge of the ``game paradox'', which translates into the need to balance playful exploration and an orientation towards computational practices during the game.

In addition to games, other resources are used to address the challenge of engaging students in the classroom. In~\cite{arPapavlasopoulou201973}, the authors identified design principles that can increase children's participation in coding activities, as well as providing theoretical and practical evidence on the value of constructionism-based coding activities. The integration of Scratch with Parsons Programming Puzzles~\cite{arBender202207} can also catalyze motivation for CT, even for adults, by reducing extrinsic cognitive load and increasing learning efficiency.

Another commonly addressed challenge that can also affect participation is the learning curve for using tools \cite{arVidal-Silva2022109, arAttard202005, arHsu202238, arGamito202231, arHou202036, arSilva201990, arKourti202348, arYallihep2020116, arZampieri2020118, arParsazadeh202175}. In addition to understanding the subjects taught, students must learn to handle the technology employed in teaching. Solving this problem can be closely related to another challenge: teacher training. After all, well-prepared professionals can better support students.

Another aspect to consider is that the earlier the use of technologies is introduced into education, the better students will adapt to using digital tools. However, this faces another challenge: adapting pedagogical practices to include CT teaching and the use of digital tools in curricula \cite{arHsu202137, arKong202347, arMa202154, arPark202274, arUrquizo2021106, arBedar202001, arErsozlu202323, arStupuriene202494, arZampieri2020118, arTucker-Raymond2021103, arMechelen202357}.

\subsection{Teacher Training}

Proper teacher training is fundamental, especially in the context of CT \cite{arCutumisu201917, arMolina-Ayuso202361}. In addition to the concepts of the subjects, students will need to learn how to use digital tools. To provide adequate support, teachers must be properly trained \cite{arMontiel202164, arChekour202313}. This issue is frequently discussed in the literature, but related challenges still cause impasses and require further research~\cite{arSanchezCamacho202387}.

In~\cite{arCaeli202010}, the authors identified that many schools in Denmark are already implementing CT-related initiatives; however, their principals highlighted the lack of teacher preparation to teach this subject. It points out the importance of professional development initiatives and cultural changes within schools to effectively incorporate CT into the educational curriculum.

In~\cite{arFESSAKIS201928}, the study highlights that Greek Computer Science teachers often view CT as merely technical, overlooking cognitive and social skills, indicating the need for clearer understanding.

In \cite{arMorze202266}, the authors reported challenges faced in the Ukrainian context, such as teachers' lack of understanding of CT and their reluctance to incorporate it into teaching, along with the scarcity of specific educational resources. In~\cite{arSaez-Lopez202085}, Spanish authors also reinforce the importance of teaching CT in the training of future teachers and point out the need for greater integration of visual programming tools in their training curricula. In line with this, in~\cite{arTikva202199}, after conducting a comprehensive literature review, the authors suggest that the proper use of tools, combined with effective learning strategies and teacher training, can promote the development of CT in students from different educational contexts.

In~\cite{arTucker-Raymond2021103}, the main contributions of the research include identifying strategies and resources used by teachers to support the computational problem solving of students. Among the challenges faced, the authors highlighted the lack of prior programming experience of teachers and the need to adapt their pedagogical practices.

In~\cite{arMonjelat202063}, the research results suggested that integrating digital tools such as Scratch into teacher professional development courses encouraged deeper reflection on the programming process and underlying concepts, providing a differentiated learning strategy. In~\cite{arKong202347, arStupuriene202494}, the authors emphasized the importance of investing in continuous high-quality professional development programs for teachers, with the aim of improving their knowledge in Computer Science content and pedagogical skills. They also recommended promoting a school culture that encourages collaboration and mutual support among teachers. This collaborative approach among teaching professionals is also advocated in~\cite{arUng2022105}. In~\cite{arMonjelat201962}, the research results highlighted the importance of approaching programming as a process close to the reality of teachers, emphasizing the relevance of mediated and situated educational practices.

Trained teachers can contribute to the development of updated pedagogical practices to incorporate digital technologies into CT teaching. In~\cite{arBroley202308, arStupuriene202494}, the authors point out the need for collaboration between educators and specialists to develop clear and comprehensive guidelines for integrating Computer Science into the elementary school curriculum, as well as choosing appropriate tools for teaching. Finally, it is important to note that there is considerable difficulty in measuring the CT skills of students. This obstacle can further complicate formulating adequate proposals for the innovations that courses need to adapt CT and digital tools in their curricula.

\subsection{Computational Thinking Assessment}

The evaluation of the effectiveness of CT education on student learning is a topic of various discussions, and the lack of widely accepted tools to measure advances in students' skills represents a significant gap \cite{arSanchezCamacho202387, arRich202280, arBedar202001, arMontiel202164, arAltanis201902, arZhang2019120, arMukasheva202168, arFagerlund202127, arChai202112, arPanskyi201971, arTang202097}. To promote a more comprehensive and effective understanding of CT, it is important to have assessments adapted for different competencies and educational levels, with evidence of reliability and validity reported when designing and evaluating such instruments.

The study conducted in~\cite{arMcCormick202256} reveals that most of the evaluated articles adopted pre- and post-test research designs to assess the impact of teaching and tool use. According to the authors, this methodology is insufficient to evaluate CT comprehension. In fact, this finding is a widely discussed issue in the literature. In \cite{arSanchezCamacho202387, arRich202280, arBedar202001, arMontiel202164, arAltanis201902, arZhang2019120, arMukasheva202168, arFagerlund202127, arChai202112, arPanskyi201971, arTang202097}, the authors also comment on this issue, but without a widely accepted solution by the community. The conclusion is that the development of pedagogically significant tools for CT assessment in programming projects and processes for students remains an existing gap. Despite some already available proposals, there has not been widespread adoption.

In~\cite{arChai202112}, for example, the main contributions include the proposition of a new Dynamic Weighted Evaluation System (DWES). The results indicate that DWES was able to provide more detailed and flexible assessments compared to existing tools like Dr. Scratch, highlighting the advantages of each type of project and improving their overall evaluations. However, the authors still mention the need to develop more comprehensive and detailed evaluation criteria for the Scratch context as a challenge.

In~\cite{arTsai2021102}, the authors developed and validated a tool called the Computational Thinking Scale (CTS), designed to assess students' CT skills. It was created based on a conceptual framework that highlights five dimensions of CT: abstraction, decomposition, algorithmic thinking, evaluation, and generalization. CTS was developed through an exploratory factor analysis approach, using statistical methods to validate the scale. This tool was designed not only to assess specific computer programming skills, but also to capture students' general tendencies in applying CT in various problem-solving contexts. Thus, CTS can be applied in diverse learning environments, regardless of the presence of programming tasks, providing a comprehensive view of students' cognitive skills related to CT. However, it may not cover all the competencies that one might want to analyze. 

For example, in~\cite{arWu2023113}, the authors wanted to measure characteristics such as motivation, anxiety, and confidence before and after the course (social and emotional competencies). These factors are not evaluated by the CTS. Therefore, the authors assessed students through specific questionnaires and, to track student progress, developed a tracking system that recorded their operations and behaviors during programming tasks. However, this proposal, like the CTS, does not encompass all the competencies that can signal improvements in CT knowledge.

In~\cite{arELOY202122}, one of the main contributions of the study was to propose a data-driven approach to evaluating CT skills using the automatic analysis of the computational artifacts of learners. The research methodology involved collecting and analyzing data from projects created by course participants, using algorithms and Python programming to extract and calculate coefficients that represent CT concepts. However, this proposal was formulated only for the programming knowledge area. In~\cite{arAltanis201902}, the authors proposed a multifaceted framework to evaluate student performance, but with specific utility throughout the game creation process, covering skills in game analysis and design, CT, programming, and spatial thinking.

\subsection{Computational Thinking vs. Mathematics}

In the literature, it is common to find articles emphasizing the significant synergistic potential between CT and Mathematics \cite{arRodriguez-Benito202082, arMolina-Ayuso202260}. Some studies have been conducted to explore this fact \cite{arGokce202332, arMolina-Ayuso202361, arKilhamn202246}, but gaps remain.

In~\cite{arErsozlu202323}, the authors comprehensively analyze the effective use of Scratch for teaching programming and solving mathematical problems, especially in educational contexts. Among the research challenges mentioned is the need for effective integration of CT into the mathematics curriculum. In~\cite{arOzcan202170} and \cite{arXie2023114}, the authors also use Scratch and highlight the importance of further investigating how mathematics education can affect CT and how programming education contributes to the development of computational skills beyond mathematical training.

In~\cite{arRodriguez-Martinez202083}, the research aimed to investigate the influence of Scratch on both the acquisition of mathematical concepts and the development of CT. The authors highlighted the scarcity of studies that address how programming activities affect learning in specific areas of mathematics. The results revealed a significant increase in the CT levels of the students, as well as a statistically significant improvement in mathematical problem solving, indicating the potential of Scratch to promote learning in both areas. However, they recognized the need for more comprehensive longitudinal research and the consideration of additional variables for a deeper understanding of the impact of programming activities on mathematics education.

In~\cite{arSantos202389, arSilva202291}, the need to include digital tools such as Scratch, MIT App Inventor, and Tinkercad in mathematics teacher training is addressed so that future teachers can use them effectively in their pedagogical practices, making mathematics teaching more meaningful, interactive, playful, and creative. In this context, in~\cite{arValentine2018107}, the authors used an exploratory and experimental teaching approach (tinkering) together with the Logo programming language, in order to engage and prepare future elementary school teachers. The conclusions highlighted the effectiveness of the logo tinkering approach in promoting a deeper understanding of mathematical concepts and a connection between mathematics and computational literacy among future teachers.

\subsection{Other Interesting Research Opportunities}

In~\cite{arMladenovic202159}, the authors use Python Tutor in programming education. Despite the advantages offered by the tool, they concluded that excessive use of visualization tools could hinder the retention of learned information. In~\cite{arCalandra202111}, the authors point out that the duration of the course may be a factor that favors the retention of the learned information. According to them, students need prolonged contact with programming practices. Although these studies have made relevant contributions, the need for an in-depth analysis of the optimal balance for using tools like Python Tutor in teaching Programming and CT remains open.

In~\cite{arVanicek2022108}, the authors used Blockly to teach the concept of loops with a fixed number of repetitions to students in 7th and 8th grade. They then used the results of the tasks completed by these students to identify misconceptions about what was taught. An interesting research could be conducted by analyzing misconceptions acquired by students in the same age group using different digital tools to learn about the same topics.

Another interesting visual programming tool discussed a few times in the literature in the context of CT is MIT App Inventor. In~\cite{arRochadiani202381}, the authors highlight the importance of interdisciplinarity combined with the use of the tool to promote CT. However, this interdisciplinarity is a challenging objective in curriculum design~\cite{arHsu202137}. The importance of more interdisciplinary approaches is also reinforced in \cite{arZampieri2020118, arWang2022110, arSjoberg202093}, which concluded that the design of activities that involve multiple disciplines has a more significant impact on the development of CT of students. Therefore, proposals for teaching CT in this context, which can be assisted by the use of MIT App Inventor and other digital tools, are welcome. In~\cite{arDeng202018}, the authors use Pencil Code and reiterate the importance of integrating CT into interdisciplinary teaching practices, suggesting that future studies should explore these issues more deeply.

In~\cite{arXing2021115}, the authors analyze codes obtained from an online programming community. Members of this community used Scratch and shared projects with each other. The study's conclusions highlight the complexity of the influence of remixing on CT, suggesting that its excessive practice can benefit and hinder the development of CT skills. The need for additional controlled experiments is suggested to better understand the impact of remixing on CT. Tools that facilitate the sharing of code or activity solution, such as Scratch, Google Colab, MIT App Inventor, Code.org, and Tinkercad, are good candidates for these experiments.

In~\cite{arMontiel202164}, authors identify the need for proposals to incorporate Scratch more broadly into school curricula. In~\cite{arPerez-Jorge202277}, the need for additional studies focused on Scratch and MIT App Inventor is commented on in the university context. In addition to Scratch, all other tools presented in Table~\ref{tab:tools_overview} are also validated and show interesting results in the CT context. Therefore, the scarcity of publications considering them reveals an underexplored scope of research. In fact, in~\cite{arFidai202029}, the authors claim that a more empirical analysis and the proposal of best practices for the use of open source tools in CT education are still open.

In~\cite{arKale202142, arKale202343}, the authors identified the need to expand lesson plans for CT courses using the Code.org tool. According to~\cite{arKale202142}, the lesson plans obtained from the tool emphasize certain skills, such as problem solving, but others equally important, such as self-regulation, analysis, and abstraction, are less addressed. In~\cite{arKale202343}, the problem is confirmed, with some intersections and additions. The authors noted that there is a prevalence of pattern recognition and algorithms/automation skills in the analyzed lesson plans. However, a significant gap was identified in the way of addressing decomposition and abstraction skills. This problem is not exclusive to Code.org's lesson plans. In~\cite{arMcCormick202256}, the authors conducted a comprehensive review of CT experiences in children with PK, exploring various types of tools used in the context. The analysis revealed a lack of emphasis on more complex CT practices, such as abstraction and modularization. This discussion highlights the gaps in the larger proposals for CT education, addressing all relevant competencies in a balanced manner.

\section{Concluding Remarks} \label{secConcluding}

This article provided a comprehensive analysis of the global landscape of Computational Thinking (CT) education, focusing on the integration of digital tools in teaching CT in various educational contexts. It reviewed the inclusion of CT in school curricula worldwide, categorized digital tools into groups such as visual programming, textual programming, electronic games, modeling, and simulation, and evaluated their usage in different educational settings. The research also identified key CT competencies enhanced through these tools and highlighted the challenges faced in implementing digital tools for CT development.

The primary objective of this study was to investigate and organize the state-of-the-art in the use of digital tools to teach CT. Specific objectives included analyzing the global presence and integration of CT in school curricula, categorizing the various digital tools used in teaching CT, examining their application in different educational contexts, and identifying the competencies developed with these tools. The study also aimed to identify challenges and gaps in the implementation of CT education, such as infrastructure inadequacies, difficulties in tool usability, teacher training needs, and adaptation of pedagogical practices.

The contributions of this research are threefold. First, it offers the reader a global overview of the inclusion of CT in school curricula, providing a broad perspective on how different countries are approaching this educational challenge. Second, it provides a comprehensive understanding of the various digital tools available for teaching CT, helping educators and policymakers select the appropriate tools for specific educational contexts. Third, the study offers a clearer and more organized view of the CT competencies that have been enhanced with these tools, helping to understand the skills developed through CT education.

Looking ahead, there are several areas for future research and exploration. One key area is the development of standardized assessment methods to evaluate CT competencies in different educational contexts. In addition, more research is needed to explore the long-term impacts of CT education on students' academic and career outcomes. Investigating the effectiveness of various teacher training programs in preparing educators to teach CT is another critical area for future work. Lastly, more studies are needed to explore the integration of emerging technologies, such as artificial intelligence and machine learning, into CT education and their potential to enhance learning outcomes.

\begin{acks}
This study was financed in part by the Postdoctoral Researcher Program (PPPD) at UNICAMP and CNPq (316067/2023-7).
\end{acks}

\bibliographystyle{ACM-Reference-Format}
\bibliography{refs}

\end{document}